\documentclass[twocolumn]{article}
\usepackage{arxiv}
\usepackage[utf8]{inputenc} 
\usepackage[T1]{fontenc}    
\usepackage{hyperref}       
\usepackage{url}            
\usepackage{booktabs}       
\usepackage{amsfonts}       
\usepackage{nicefrac}       
\usepackage{microtype}      
\usepackage{lipsum}
\usepackage{fancyhdr}       
\usepackage{graphicx}       
\usepackage[export]{adjustbox} 
\usepackage{stfloats}       
\fnbelowfloat               
\graphicspath{{media/}}     
\usepackage[labelfont=bf,justification=raggedright,singlelinecheck=false]{caption} 

\usepackage{caption}
\title{Physics-Informed Sinusoidal Representation Networks for Damage Identification in Ultrasonic Guided Wave Structural Health Monitoring}

\author{
  Lukas Jilke\thanks{German Aerospace Center (DLR), Institute of Maintenance, Repair and Overhaul, Aachen-Würselen, Germany} \\
  \And
  Caspar Wasle\thanks{German Aerospace Center (DLR), Institute of Maintenance, Repair and Overhaul, Hamburg, Germany} \\
  \And
  Norbert Hosters\thanks{RWTH Aachen University, Chair for Computational Analysis of Technical Systems, Aachen, Germany} \\
  \AND
  Florian Raddatz\footnotemark[2] \\
  \And
  Marek Behr\footnotemark[3] \\
  \And
  Gerko Wende\footnotemark[2] \\
}

\begin{document}

\begin{abstract}
Ultrasonic guided wave (UGW) inspection has become an effective method in structural health monitoring (SHM). While machine learning has enhanced damage detection capabilities, purely data-driven models remain constrained by their reliance on extensive labeled datasets and limited physical interpretability. Standard multi-layer perceptrons (MLPs) further suffer from spectral bias (F-Principle) in high-frequency regimes, prioritizing low-frequency components while failing to resolve the high-frequency oscillations critical for UGW reconstruction. Physics-Informed Sinusoidal Representation Networks (PI-SIRENs) address these limitations by embedding governing partial differential equations directly into the training process. Building upon SIRENs, which utilize periodic activation functions to inherently capture high-frequency spectral content, the proposed framework is designed to operate effectively in sparse ultrasonic measurement regimes. This work introduces a frequency-domain PI-SIREN framework specifically tailored for the Mindlin--Reissner plate theory, embedding the coupled elastodynamic equations to reconstruct ultrasonic wavefields without temporal discretization. The model parameters are optimized using a hybrid strategy comprising three sequential phases: first-order Adam optimization, residual-adaptive refinement (RAR) of the collocation set and quasi-Newton L-BFGS. Numerical and experimental evaluations confirm that the trained model serves as a reliable physics-consistent baseline derived from sparse data, functioning as a computationally efficient surrogate for damage identification in SHM and non-destructive testing (NDT) applications. Integrating the trained surrogate into a proposed damage identification pipeline detects and localizes a milled notch from sparse experimental measurements with a localization error of $1\,\%$ of the plate diagonal. In contrast, purely data-driven benchmarks and MLPs utilizing hyperbolic tangent activation functions, with and without Fourier feature input embedding, are unable to identify damage due to higher reconstruction errors. 
\end{abstract}


\correspondingauthor{$^*$Lukas Jilke}{Lukas.Jilke@DLR.de}

\citationnote{\textbf{Preprint}: This paper has been submitted for publication and is currently under peer review.}


\maketitle
\section{Introduction}
\label{sec:Introduction}
Ultrasonic guided wave (UGW) inspection has emerged as a central technique in structural health monitoring (SHM) applications, particularly for thin-walled aerospace structures, where the early detection of material degradation is essential~\cite{Jilke.2025}. UGWs are elastic waves propagating within bounded media, characterized by low attenuation over long distances and high sensitivity to localized perturbations such as cracks, delaminations or corrosion~\cite{Willberg.2015, Abbas.2018}. The increasing complexity of structural systems and the rising costs of maintenance, repair and overhaul (MRO) activities require the development of rapid, reliable and quantitative damage identification methods~\cite{Jilke.2023}. Conventional physics-based simulations, such as the finite element method (FEM) and boundary element method (BEM), provide high-fidelity wavefield simulations but remain computationally expensive for real-time monitoring or inverse analyses~\cite{Huang.2016}. In contrast, purely data-driven machine learning approaches provide rapid training but require extensive labeled datasets and often lack physical interpretability, limiting their robustness in sparse-measurement regimes~\cite{Nerlikar.2024, Rautela.2021, Yang.2023}. Hybrid modeling frameworks that embed physical laws into data-driven architectures have therefore gained traction, providing a balance between computational efficiency and physical consistency. Physics-informed neural networks (PINNs) have emerged as hybrid frameworks that incorporate physical knowledge during the training phase~\cite{Raissi.2019}. By enforcing partial differential equations as constraints within the loss function, PINNs provide essential regularization that enables reliable predictions in regimes with limited observational data~\cite{Travnikova.2025}. In the context of SHM and ultrasonic non-destructive testing (NDT) applications, several studies have applied PINNs to damage inference in various tasks. Al-Adly et al.~\cite{AlAdly.2024} demonstrate the application of a PINN model for quasi-static plate bending under Kirchhoff--Love (KL) assumptions, focusing on state estimation and virtual sensing. The authors show that this approach can accurately predict unmeasured structural deflections and internal bending moments even under variable loading and noisy data conditions. Since this approach operates strictly within a spatial and quasi-static regime, it does not account for temporal dynamics, making it inapplicable to real-time tracking, dynamic anomaly detection or ultrasonic wave-based damage identification. Zhou and Xu~\cite{Zhou.2024} introduce a baseline-free damage identification method for plate structures designed to isolate localized defects using UGW without requiring experimental historical baselines. Their approach applies multi-layer perceptron (MLP)-based PINNs governed by the KL plate theory to UGW data, constructing a pseudo-pristine model for damage identification based on a two-dimensional Teager energy operator metric. However, the framework remains constrained to lower-frequency excitations and leaves boundary conditions explicitly unenforced, which requires higher-density spatial measurement data and computationally expensive model retraining for every encountered measurement. Song et al.~\cite{Song.2026} pursue a different baseline-free approach, training a physics-guided multiscale network on wavelet coefficients of time-domain Lamb wave signals to eliminate boundary reflections and extract scattered-wave arrival times without explicit wavefield reconstruction. Shukla et al.~\cite{Shukla.2020} localize surface-breaking cracks in aluminum plates by training a PINN with adaptive activation functions on ultrasonic laser vibrometry data. The framework relies on the 2D scalar acoustic wave equation to model surface displacements and is primarily limited to modeling the behavior of Rayleigh waves in thick plates. The model reconstructs local sound-speed maps that indicate defect geometries. However, it lacks both the elastodynamic formulations required to capture dispersive guided waves in thin plates and the generalization capability to handle new inspection scenarios without model retraining. Huang et al.~\cite{Huang.2025} achieve reliable micro-crack localization and depth characterization by mapping equivalent velocity reductions from out-of-plane surface wave displacements, demonstrating robust performance with only 50\% of the required training data. However, the framework is restricted to an idealized one-dimensional scan path and employs a decoupled empirical linear regression to infer crack depth rather than performing direct physical parametric inversion. Additionally, the model requires re-optimization for new measurement configurations. Kulkarni and Sabato~\cite{Kulkarni.2026} combine a variational autoencoder with physics-informed constraints to expand sparse thermal measurements to full-field temperature distributions for defect detection on metal plates. 

A fundamental limitation of conventional PINN architectures is spectral bias, also referred to as the F-Principle~\cite{Xu.2025c, Khodakarami.2026}. Standard MLPs tend to prioritize learning the low-frequency components of the target function, making them prone to converging to suboptimal local minima~\cite{Rahaman.2018}. In UGW modeling, this bias manifests as smoothed wavefronts, attenuated scattering signatures and failure to capture phase-dependent interference patterns essential for damage quantification. To address spectral bias, modifications to network architectures have been explored in various studies. Sitzmann et al.~\cite{Sitzmann.2020} introduce Sinusoidal Representation Networks (SIRENs), which employ periodic activation functions that inherently support multi-scale frequency representation. Recent applications in wave propagation analysis have further improved spectral resolution through multi-scale Fourier feature embeddings~\cite{Song.2022, Ding.2024, Schoder.2025}. By embedding Fourier features into a dual-network PINN architecture, Chen et al.~\cite{Chen.2026} mitigate spectral bias to invert high-resolution phase velocity drops from sparse laser ultrasonic data and characterize densely distributed near-surface defects. However, the proposed framework leaves the full elastodynamic equations unexamined, remains restricted to idealized cylindrical cavity geometries in isotropic media and requires a separate model re-optimization cycle for each unique measurement wavefield. Recent work by Yang et al.~\cite{Yang.2026} demonstrates that defect-parameterized PINN surrogates can effectively reconstruct scattered wavefields from sparse laser ultrasonic measurements, enabling fast inverse characterization of crack geometry. However, their framework relies on time-domain training and empirical transition steepness tuning, which complicates deployment in practical SHM workflows where rapid baseline generation and standardized damage identification pipelines are required. Despite these advances, the majority of PINN formulations for UGWs are formulated in the time domain, which adds a temporal dimension and introduces known causality issues during training~\cite{Wang.2024}. A dedicated frequency-domain framework operating on the coupled spatial fields remains underexplored in UGW analysis. Frequency-domain PINNs offer distinct advantages for UGW analysis by eliminating temporal derivatives and initial condition dependencies, reducing the problem domain to a stationary boundary value problem~\cite{Jilke.2026}. At typical UGW excitation frequencies, shear deformation and rotational inertia cannot be neglected. The Mindlin--Reissner (MR) plate theory~\cite{Mindlin.1951, Reissner.1945} provides the kinematic framework to capture these effects through coupled transverse displacement and sectional rotation fields. 

The present study introduces a frequency-domain PI-SIREN architecture formulated for the MR plate theory to address these limitations. The model embeds the coupled elastodynamic equations as constraints within the training loss of a SIREN-based neural network, enabling the reconstruction of steady-state guided wavefields without temporal discretization or explicit initial condition constraints. By employing periodic activations, the architecture counteracts spectral bias and thereby resolves the high-frequency ultrasonic oscillations that standard MLPs utilizing $\tanh$ activation functions fail to capture. The network parameters are optimized using a hybrid three-phase strategy: first-order Adam~\cite{Kingma.2014}, residual-adaptive refinement (RAR)~\cite{Lu.2021b} and quasi-Newton L-BFGS~\cite{Liu.1989, Bollapragada.2018}. The primary contributions of this work are:
\begin{itemize}
    \item A frequency-domain PI-SIREN framework formulated for the coupled MR plate equations, enabling steady-state wavefield reconstruction from sparse measurements without temporal discretization.
    \item Numerical and experimental validation demonstrating that the trained model functions as a physics-consistent baseline surrogate, outperforming data-only and no-RAR variants in ablation studies and $\tanh$ MLP baselines with and without Fourier feature embedding.
    \item A two-stage damage identification pipeline integrating the surrogate model with coarse residual scanning and adaptive local refinement, achieving high localization accuracy on experimental ultrasonic wavefield measurements.
\end{itemize}

\section{Problem Formulation}
\label{sec:Problem}
Reliable damage detection and localization in UGW SHM require establishing a physically consistent pristine baseline for comparison against measured structural responses~\cite{Rautela.2021}. While Rytter's hierarchy~\cite{Rytter.1993} establishes the theoretical steps from damage detection to severity quantification, practical deployments face a fundamental challenge: acquiring high-fidelity baseline measurements across all operational conditions is rarely feasible. Finite element (FE) simulations can provide such baselines but are computationally expensive for repeated evaluations. Furthermore, experimental measurements are inherently sparse due to limited sensor placement, making purely data-driven approaches insufficient for reconstructing the full wavefield without physical regularization. To address these limitations, this work formulates the forward wave propagation problem as a physics-informed regression task. The structure under consideration is a homogeneous, isotropic, linearly elastic plate with lateral dimensions $L$ and thickness $h$, excited with a five-cycle Hann-windowed burst at a driving frequency of $100\,\mathrm{kHz}$. At this frequency, shear deformation and rotational inertia effects become significant, necessitating the use of first-order shear deformation theory to accurately capture the dispersive UGW modes~\cite{Mindlin.1951, Lints.2024}. The objective is to learn a continuous, differentiable surrogate model $\mathcal{N}_{\boldsymbol{\theta}} : (\bar{x}, \bar{y})\mapsto\hat{\boldsymbol{w}}$ that predicts the steady-state transverse displacement wavefield at arbitrary spatial coordinates from sparse measurement data. The parameters $\boldsymbol{\theta}$ are obtained by minimizing a composite loss function that enforces the physical constraints while ensuring data fidelity.

Training such surrogates introduces significant computational challenges. First, standard neural networks suffer from spectral bias (the F-Principle), inherently struggling to approximate high-frequency oscillatory solutions characteristic of wave propagation~\cite{Rahaman.2018}. Second, the MR formulation involves three coupled second-order partial differential equations (PDEs), which can yield an ill-conditioned loss landscape. These factors combine into a highly non-convex optimization problem prone to convergence to suboptimal local minima. To overcome these challenges, this work employs a network architecture with sinusoidal activation functions that capture the high-frequency oscillations and stabilize gradient propagation during training~\cite{Sitzmann.2020}.

By embedding the coupled elastodynamic equations of MR plate theory into the network training, the framework enables accurate wavefield reconstruction from sparse inputs while reducing measurement effort. The resulting physics-consistent baseline serves as a computationally efficient alternative to repeated FE simulations and functions as the reference state in the proposed two-stage damage identification pipeline, in which localized scattering signatures are isolated through adaptive comparison against the pristine-state surrogate prediction.

\section{Governing Equations}
\label{sec:governing_equations}
The propagation of elastic waves within a continuous, homogeneous and isotropic solid is governed by the Navier--Cauchy equations for elastodynamics. Assuming steady-state time-harmonic excitation of the form $e^{\mathrm{i}\omega t}$, temporal derivatives transform as $\ddot{\boldsymbol{u}} \rightarrow -\omega^2 \hat{\boldsymbol{u}}$, reducing the problem to the frequency domain. Derived from the conservation of linear momentum and Hooke's law, the general three-dimensional equation of motion in the absence of volumetric body forces is expressed as~\cite{Rose.2014}:
\begin{equation}
    -\rho \omega^2 \hat{\boldsymbol{u}} = (\lambda + \mu)\,\nabla(\nabla \cdot \hat{\boldsymbol{u}}) + \mu\,\nabla^2 \hat{\boldsymbol{u}},
    \label{eq:navier_lame_freq}
\end{equation}
where $\omega = 2\pi f$ is the angular frequency, $\hat{\boldsymbol{u}}(\boldsymbol{x},\omega) \in \mathbb{C}^3$ denotes the complex displacement amplitude, with $\boldsymbol{x} = (x, y, z)$ as the spatial coordinate vector, $\rho$ is the mass density and $\lambda$ and $\mu$ are the Lam\'{e} constants defined as:
\begin{equation}
    \lambda = \frac{E\nu}{(1+\nu)(1-2\nu)}, \qquad \mu = \frac{E}{2(1+\nu)}.
\end{equation}
The formulation omits external forces, since the present study focuses on the steady-state response to the excitation, resulting in a source-free homogeneous system of equations. Solving the full three-dimensional elastodynamic equations~\eqref{eq:navier_lame_freq} is computationally expensive. Plate theories reduce the computational effort for wave propagation in plate-like structures where the thickness $h$ is significantly smaller than the lateral dimensions $L$. To reduce the problem to two dimensions, the system of equations is simplified by means of the MR first-order shear deformation theory. This theory relaxes the strict normality constraint of classical thin-plate formulations by allowing cross-sections to shear. The fundamental kinematic assumption states that straight lines perpendicular to the undeformed mid-surface remain straight and inextensible during deformation but are not required to remain perpendicular to the deformed mid-surface~\cite{Mindlin.1951, Reissner.1945}. This introduces two independent rotational degrees of freedom, $\hat{\psi}_x$ and $\hat{\psi}_y$, alongside the transverse displacement $\hat{w}$. The three-dimensional displacement vector field in $(x,y,z)$-direction $\hat{\boldsymbol{u}} = (\hat{u},\hat{v},\hat{w})^\top$ is approximated as:
\begin{align}
    \hat{u}(x,y,z,\omega) &= z\,\hat{\psi}_x(x,y,\omega), \\
    \hat{v}(x,y,z,\omega) &= z\,\hat{\psi}_y(x,y,\omega), \\
    \hat{w}(x,y,z,\omega) &= \hat{w}(x,y,\omega),
\end{align}
where $z \in [-h/2,\, h/2]$ is the through-thickness coordinate. Substituting this kinematic Ansatz into the three-dimensional strain-displacement and constitutive relations and integrating the resulting stresses over the plate thickness yields the steady-state equations of motion. Expressed in terms of complex stress resultants, bending moments $\hat{M}_{(\cdot)}$ and transverse shear forces $\hat{Q}_{(\cdot)}$, the three coupled equilibrium equations are given by~\cite{Lints.2024}:
\begin{align}
    \frac{\partial \hat{M}_{xx}}{\partial x} + \frac{\partial \hat{M}_{xy}}{\partial y} - \hat{Q}_x + I_2\,\omega^2 \hat{\psi}_x &= 0, \label{eq:MR_eq_1_freq} \\
    \frac{\partial \hat{M}_{xy}}{\partial x} + \frac{\partial \hat{M}_{yy}}{\partial y} - \hat{Q}_y + I_2\,\omega^2 \hat{\psi}_y &= 0, \label{eq:MR_eq_2_freq} \\
    \frac{\partial \hat{Q}_x}{\partial x} + \frac{\partial \hat{Q}_y}{\partial y} + \rho h\,\omega^2 \hat{w} &= 0, \label{eq:MR_eq_3_freq}
\end{align}
where $I_2 = \rho h^3/12$ is the rotary inertia per unit area. The constitutive relations linking the stress resultants to the kinematic variables are:
\begin{align}
    \hat{M}_{xx} &= D\left(\frac{\partial \hat{\psi}_x}{\partial x} + \nu\,\frac{\partial \hat{\psi}_y}{\partial y}\right), \label{eq:M_xx} \\
    \hat{M}_{yy} &= D\left(\frac{\partial \hat{\psi}_y}{\partial y} + \nu\,\frac{\partial \hat{\psi}_x}{\partial x}\right), \\
    \hat{M}_{xy} &= D\,\frac{1-\nu}{2}\left(\frac{\partial \hat{\psi}_x}{\partial y} + \frac{\partial \hat{\psi}_y}{\partial x}\right), \\
    \hat{Q}_x    &= \kappa G h\left(\hat{\psi}_x + \frac{\partial \hat{w}}{\partial x}\right), \\
    \hat{Q}_y    &= \kappa G h\left(\hat{\psi}_y + \frac{\partial \hat{w}}{\partial y}\right), \label{eq:Qy}
\end{align}
where $D = Eh^3/[12(1-\nu^2)]$ is the flexural rigidity, $G = E/[2(1+\nu)]$ is the shear modulus and $\kappa = 5/6$ is the shear correction factor for rectangular cross-sections. The sign convention follows from the Ansatz $\hat{u} = z\hat{\psi}_x$, for which the transverse shear $\hat{Q}_x$ vanishes in the KL limit $\hat{\psi}_x \to -\partial \hat{w}/\partial x$, consistent with the classical constraint that normals remain perpendicular to the deformed mid-surface~\cite{Mindlin.1951}.
At the plate surface $z = h/2$, the linear through-thickness displacement field relates the in-plane surface displacements $\hat{u}_{\text{surf}}$ and $\hat{v}_{\text{surf}}$ to the rotations as:
\begin{equation}
    \hat{\psi}_x = \frac{2}{h}\,\hat{u}_{\text{surf}}, \qquad \text{and} \qquad \hat{\psi}_y = \frac{2}{h}\,\hat{v}_{\text{surf}}.
\label{eq:surface_displacements}
\end{equation}

\section{Methodology}
\label{sec:Methodology}
The proposed frequency-domain PI-SIREN framework, designed to solve the coupled elastodynamic equations of MR plate theory, is detailed in the following subsections. These subsections present the neural network architecture for high-frequency spectral representation, the formulation of the tailored physics-informed loss function and the proposed hybrid optimization strategy integrating Adam, RAR and L-BFGS fine-tuning. 

\subsection{PI-SIREN Architecture}
\label{subsec:PI-SIREN-Model}
The surrogate model is constructed as a PINN based on the SIREN architecture. Unlike conventional architectures that utilize standard non-periodic activation functions such as hyperbolic tangent, SIRENs employ periodic sinusoidal activations of the form $\phi(\xi) = \sin(\omega_0 \xi)$, which are particularly effective for learning the high-frequency oscillations and complex spectral representations inherent in UGW propagation~\cite{Sitzmann.2020}. Here $\xi$ denotes the pre-activation input, defined as the affine transformation of the outputs from the previous layer. Conventional MLPs with smooth, non-periodic activations such as $\tanh$ fail to represent high-frequency information and, more critically, its higher-order derivatives, even under dense supervision~\cite{Sitzmann.2020}. The frequency-domain formulation eliminates the temporal dimension, reducing the network input to normalized two-dimensional spatial coordinates $(\bar{x}, \bar{y}) \in [-1, 1]^2$. The network outputs the three complex surface-displacement amplitudes $(\hat{\bar{u}},\, \hat{\bar{v}},\, \hat{\bar{w}})$ at $f_0$, normalized by the coordinate map and field scales, each decomposed into its real and imaginary parts, yielding a six-dimensional output vector:
\begin{equation}
    \boldsymbol{x}_{\text{out}} 
    = \begin{bmatrix}
        \mathrm{Re}(\hat{\bar{u}}) \\
        \mathrm{Im}(\hat{\bar{u}}) \\
        \mathrm{Re}(\hat{\bar{v}}) \\
        \mathrm{Im}(\hat{\bar{v}}) \\
        \mathrm{Re}(\hat{\bar{w}}) \\
        \mathrm{Im}(\hat{\bar{w}}) 
      \end{bmatrix}.
    \label{eq:network_output}
\end{equation}
The rotational fields $\hat{\psi}_x$ and $\hat{\psi}_y$ required by the MR plate theory are derived analytically from the predicted displacements using first-order shear deformation kinematics according to Eq.~\eqref{eq:surface_displacements}. During training, only the transverse displacement $\hat{\bar{w}}$ is constrained by measurement data. The in-plane displacements and the derived rotational fields are learned implicitly through the implemented governing PDEs and boundary conditions (BCs). This parameterization matches the quantity that is directly observable in the experiments.

\begin{figure*}[htbp]
\centering
\includegraphics[width=0.95\textwidth]{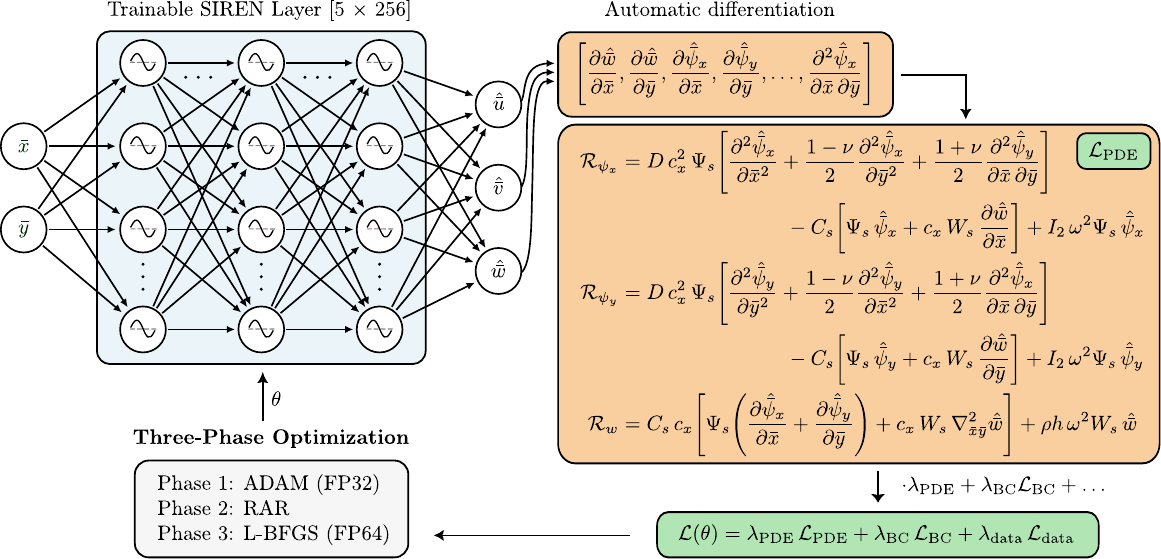}
\caption{Schematic representation of the frequency-domain PI-SIREN architecture for solving the normalized MR plate theory equations.}
\label{fig:PI-SIREN-Architecture}
\end{figure*}

The model employs a fully connected neural network with $N_L = 5$ hidden layers, each containing $N = 256$ neurons per layer. Automatic differentiation (AD) is used to compute the spatial derivatives required for the PDE residuals. To preserve activation variance and ensure stable gradient flow across the network, the weights are initialized according to Sitzmann et al.~\cite{Sitzmann.2020}. Specifically, the first layer uses $\mathcal{U}(-1/n_{in}, 1/n_{in})$, while all subsequent layers use $\mathcal{U}(-\sqrt{6/n_{in}}/\omega_0, \sqrt{6/n_{in}}/\omega_0)$, where $n_{in}$ denotes the number of input neurons to each layer. The SIREN frequency hyperparameter is set to $\omega_0 = 30$. Figure~\ref{fig:PI-SIREN-Architecture} provides a schematic representation of the proposed PI-SIREN architecture. 

\subsection{Loss Function}
\label{subsec:loss_function}
The network parameters $\boldsymbol{\theta}$ are optimized by minimizing the composite loss function that enforces the governing physics, boundary conditions and measurement data. The overall objective function to be minimized is defined as: 
\begin{equation}
    \mathcal{L}_{\text{total}}(\boldsymbol{\theta}) = \lambda_{\text{PDE}}\,\mathcal{L}_{\text{PDE}} + \lambda_{\text{BC}}\,\mathcal{L}_{\text{BC}} + \lambda_{\text{data}}\,\mathcal{L}_{\text{data}},
    \label{eq:total_loss}
\end{equation}
where $\lambda_{\text{PDE}}$, $\lambda_{\text{BC}}$ and $\lambda_{\text{data}}$ are constant weighting coefficients that balance the relative contribution of each loss term during optimization. The PDE residual terms are formulated in normalized coordinates and scaled with field variables to ensure stable optimization. The physical domain $(x,y) \in [0,L]^2$ is mapped to the computational domain $(\bar{x},\bar{y}) \in [-1,1]^2$ by
\begin{equation}
    \bar{x} = 2\!\left(\frac{x}{L}\right) - 1, \qquad \bar{y} = 2\!\left(\frac{y}{L}\right) - 1,
\end{equation}
which introduces the chain-rule scaling factors $c_x = c_y = 2/L$ into the spatial derivatives: $\partial_x = c_x\,\partial_{\bar{x}}$ and $\partial_x^n = c_x^n\,\partial_{\bar{x}}^n$. The physical fields are scaled to $O(1)$ as $\hat{w} = W_s\hat{\bar{w}}$, $\hat{u}_{\text{surf}} = U_s\hat{\bar{u}}$ and $\hat{v}_{\text{surf}} = V_s\hat{\bar{v}}$, where $W_s$ is the maximum complex-magnitude amplitude of the transverse displacement in the reference data. For experimental data, $U_s$ and $V_s$ are not directly measurable and are set relative to $W_s$ from dispersion analysis and parametric studies as $U_s = V_s = W_s/2$. Combining the surface scaling with the kinematic relation of Eq.~\eqref{eq:surface_displacements}, the rotations scale as
\begin{equation}
    \hat{\psi}_x = \Psi_s\,\hat{\bar{\psi}}_x, \quad
    \hat{\psi}_y = \Psi_s\,\hat{\bar{\psi}}_y, \quad
    \quad \Psi_s = \frac{2\,U_s}{h},
\end{equation}
and the normalized rotations are identified with the in-plane network outputs, $\hat{\bar{\psi}}_x \equiv \hat{\bar{u}}$ and $\hat{\bar{\psi}}_y \equiv \hat{\bar{v}}$. Substituting the constitutive laws, the scaling and the normalized coordinates into Eqs.~\eqref{eq:MR_eq_1_freq}--\eqref{eq:MR_eq_3_freq} yields the PDE residuals $\mathcal{R}$:
\begin{align}
    \mathcal{R}_{\psi_x} &= D\,c_x^2\,\Psi_s\!\left[\frac{\partial^2 \hat{\bar{\psi}}_x}{\partial \bar{x}^2} + \frac{1-\nu}{2}\frac{\partial^2 \hat{\bar{\psi}}_x}{\partial \bar{y}^2} + \frac{1+\nu}{2}\frac{\partial^2 \hat{\bar{\psi}}_y}{\partial \bar{x}\,\partial \bar{y}}\right] \nonumber \\
    & - C_s\!\left[\Psi_s\,\hat{\bar{\psi}}_x + c_x\,W_s\,\frac{\partial \hat{\bar{w}}}{\partial \bar{x}}\right] + I_2\,\omega^2\Psi_s\,\hat{\bar{\psi}}_x, \label{eq:R_psix} \\
    \mathcal{R}_{\psi_y} &= D\,c_y^2\,\Psi_s\!\left[\frac{\partial^2 \hat{\bar{\psi}}_y}{\partial \bar{y}^2} + \frac{1-\nu}{2}\frac{\partial^2 \hat{\bar{\psi}}_y}{\partial \bar{x}^2} + \frac{1+\nu}{2}\frac{\partial^2 \hat{\bar{\psi}}_x}{\partial \bar{x}\,\partial \bar{y}}\right] \nonumber \\
    & - C_s\!\left[\Psi_s\,\hat{\bar{\psi}}_y + c_y\,W_s\,\frac{\partial \hat{\bar{w}}}{\partial \bar{y}}\right] + I_2\,\omega^2\Psi_s\,\hat{\bar{\psi}}_y, \label{eq:R_psiy} \\
    \mathcal{R}_{w} &= C_s\, c_x\!\left[\Psi_s\!\left(\frac{\partial \hat{\bar{\psi}}_x}{\partial \bar{x}} + \frac{\partial \hat{\bar{\psi}}_y}{\partial \bar{y}}\right) + c_x\,W_s\,\nabla^2_{\bar{x}\bar{y}}\hat{\bar{w}}\right] \nonumber \\
    & \quad + \rho h\,\omega^2 W_s\,\hat{\bar{w}}, \label{eq:R_w} 
\end{align}
where $C_s = \kappa G h$ is the transverse shear stiffness and $\nabla^2_{\bar{x}\bar{y}} = \partial^2/\partial\bar{x}^2 + \partial^2/\partial\bar{y}^2$ the two-dimensional Laplacian. The PDE loss $\mathcal{L}_{\text{PDE}}$ penalizes violations of the three coupled MR equations in Eqs.~\eqref{eq:R_psix}--\eqref{eq:R_w}, which are evaluated at $N_f$ collocation points sampled randomly across the normalized spatial domain $(\bar{x}, \bar{y}) \in [-1, 1]^2$. Defining $\mathcal{R}^{(j)} = [\mathcal{R}_{\psi_x}^{(j)}, \mathcal{R}_{\psi_y}^{(j)}, \mathcal{R}_w^{(j)}]^\top$ as the complex residual vector at the $j$-th collocation point (encompassing both real and imaginary parts), the PDE loss is formulated as:
\begin{multline}
    \mathcal{L}_{\text{PDE}} = \frac{1}{N_f} \sum_{j=1}^{N_f} \Big( 
    s_{\text{bend}}^2 \left|\mathcal{R}_{\psi_x}^{(j)}\right|^2
    + s_{\text{bend}}^2 \left|\mathcal{R}_{\psi_y}^{(j)}\right|^2 \\
    + s_{\text{vert}}^2 \left|\mathcal{R}_w^{(j)}\right|^2 
    \Big),
    \label{eq:pde_loss}
\end{multline}
where $s_{\text{vert}} = 1/\mathcal{M}_{\text{vert}}$ and $s_{\text{bend}} = 1/\mathcal{M}_{\text{bend}}$ are per-equation scaling factors, ensuring that each squared residual contributes a dimensionless term of order $\mathcal{O}(1)$ at initialization. The corresponding characteristic magnitudes $\mathcal{M}_{\text{bend}}$ and $\mathcal{M}_{\text{vert}}$ are estimated from the dominant residual terms at the excitation frequency as:
\begin{align}
    \mathcal{M}_{\text{bend}} &= \max\bigl( D c_x^2 \Psi_s,\; C_s \Psi_s, \notag \\
    &\qquad\qquad\qquad C_s c_x W_s,\; I_2\,\omega^2 \Psi_s \bigr), \\
    \mathcal{M}_{\text{vert}} &= \max\bigl(C_s c_x \Psi_s,\; C_s c_x^2 W_s,\;\rho h\,\omega^2 W_s \bigr).
    \label{eq:scaling_factors}
\end{align}
The boundary condition loss $\mathcal{L}_{\text{BC}}$ enforces free-edge conditions on all four plate edges. Specifically, for edges normal to the $\bar{x}$-direction ($\bar{x} = -1$ and $\bar{x} = 1$), the boundary residuals are defined by the vanishing of the bending moment $\hat{M}_{xx}$, twisting moment $\hat{M}_{xy}$ and transverse shear force $\hat{Q}_x$. For edges normal to the $\bar{y}$-direction ($\bar{y} = -1$ and $\bar{y} = 1$), the corresponding residuals involve $\hat{M}_{yy}$, $\hat{M}_{xy}$ and $\hat{Q}_y$:
\begin{align}
    \mathcal{R}^{x}_{M_{nn}} &= \hat{M}_{xx}, &
    \mathcal{R}^{x}_{M_{ns}} &= \hat{M}_{xy}, &
    \mathcal{R}^{x}_{Q_n} &= \hat{Q}_x, \\
    \mathcal{R}^{y}_{M_{nn}} &= \hat{M}_{yy}, &
    \mathcal{R}^{y}_{M_{ns}} &= \hat{M}_{xy}, &
    \mathcal{R}^{y}_{Q_n} &= \hat{Q}_y,
\end{align}
where the stress resultants $\hat{M}_{(\cdot)}$ and $\hat{Q}_{(\cdot)}$ are evaluated from the constitutive relations in Eqs.~\eqref{eq:M_xx}--\eqref{eq:Qy} using the normalized fields and the scaling factors defined above. The boundary loss evaluated at $N_b$ points is then defined as:
\begin{multline}
    \mathcal{L}_{\text{BC}} = \frac{1}{N_b} \sum_{k=1}^{N_b} \Big( 
    s_{M}^2 \left|\mathcal{R}_{M_{nn}}^{(k)}\right|^2 
    + s_{M}^2 \left|\mathcal{R}_{M_{ns}}^{(k)}\right|^2 \\
    + s_{Q}^2 \left|\mathcal{R}_{Q_{n}}^{(k)}\right|^2 \Big),
    \label{eq:bc_loss}
\end{multline}
where $s_{M} = 1/\mathcal{M}_{M}$ and $s_{Q} = 1/\mathcal{M}_{Q}$ are scaling factors derived from the characteristic magnitudes of bending moments and shear forces, respectively. These magnitudes are estimated based on the bending stiffness and transverse shear stiffness contributions:
\begin{align}
    \mathcal{M}_{M} &\approx D c_x \Psi_s, \\
    \mathcal{M}_{Q} &\approx C_s \max\!\left(\Psi_s, c_x W_s\right).
\end{align}
This ensures balanced gradient signals across moment and shear constraints. The global loss weights are set to $\lambda_{\text{PDE}} = 1$, $\lambda_{\text{data}} = 10$ and $\lambda_{\text{BC}} = 10$ to account for the stricter enforcement required by boundary conditions relative to the domain residuals. 

Finally, the data loss $\mathcal{L}_{\text{data}}$ ensures fidelity to the numerical and experimental benchmark by comparing the network's predictions at $N_d$ spatial locations against FE or experimental data. Data supervision is applied exclusively to the transverse displacement $\hat{\bar{w}}$. The data loss is computed as:
\begin{equation}
    \mathcal{L}_{\text{data}} = \frac{1}{N_d} \sum_{i=1}^{N_d} \left| \hat{\bar{w}}_{\text{pred}}^{(i)} - \hat{\bar{w}}_{\text{ref}}^{(i)} \right|^2,
    \label{eq:data_loss}
\end{equation}
where $\hat{\bar{w}}_{\text{pred}}$ and $\hat{\bar{w}}_{\text{ref}}$ denote the predicted and reference normalized transverse displacement vectors, respectively. 

\subsection{Optimization Strategy}
\label{subsec:Optimization_Strategy}
For the training of the proposed PI-SIREN architecture, a hybrid optimization strategy is employed. The methodology sequentially combines first-order optimization with Adam~\cite{Kingma.2014}, RAR~\cite{Lu.2021b} of the collocation set and quasi-Newton fine-tuning with L-BFGS~\cite{Liu.1989, Bollapragada.2018}. The optimization process is summarized in Algorithm~\ref{alg:FreqDomain_PINN} and detailed below.
\begin{algorithm}[htbp]
\caption{Optimization strategy}
\label{alg:FreqDomain_PINN}
\begin{algorithmic} 
    \REQUIRE Collocation budget $K$, Candidate set size $N_{\text{cand}}$, Gradient-norm tolerance $\tau$, Learning rate $\eta$, Adam iterations $N_{\text{Adam}}$, max L-BFGS iterations $M_{\max}$\vspace{1mm}
    \STATE \textbf{Initialize:} Network parameters $\boldsymbol{\theta}_0$
    \vspace{1mm}
    \STATE \textbf{Phase 1: Adam Optimization (FP32)} 
        \FOR{$n = 0, \dots, N_{\text{Adam}}-1$}
            \STATE \quad Sample coll. set $\mathcal{T}^{(n)} \sim \mathcal{U}([-1,1]^2)$, $|\mathcal{T}^{(n)}| = N_f$.
            \STATE \quad Evaluate $\mathcal{L}_{\text{total}}(\boldsymbol{\theta}_n;\, \mathcal{T}^{(n)})$, update via Adam:\vspace{1mm}
            \STATE \qquad $\boldsymbol{\theta}_{n+1} \gets \operatorname{Adam}\bigl(\boldsymbol{\theta}_n,\; \nabla_{\boldsymbol{\theta}}\mathcal{L}_{\text{total}}(\boldsymbol{\theta}_n; \mathcal{T}^{(n)}),\; \eta\bigr)$
        \ENDFOR
    \vspace{2mm}
    \STATE \textbf{Phase 2: Residual-Adaptive Refinement (RAR)}
    \STATE \quad Generate candidate set $\mathcal{S}_{\text{cand}} \subset [-1,1]^2$ of size 
    \STATE \quad $N_{\text{cand}}$, evaluate residuals $r_i = \|\mathcal{R}(\boldsymbol{\xi}_i;\boldsymbol{\theta}_{N_{\text{Adam}}})\|^2$ 
    \STATE \quad for $\boldsymbol{\xi}_i \in \mathcal{S}_{\text{cand}}$. Construct optimized collocation set:\vspace{1mm}
    \STATE \qquad $\mathcal{T}_{\text{L-BFGS}} \leftarrow \operatorname{top}_{K_{\text{hard}}}(\{r_i\}) \cup \operatorname{rand}_{K_{\text{rand}}}(\mathcal{T}^{(n)})$
    \vspace{2mm}
    \STATE \textbf{Phase 3: L-BFGS fine-tuning (FP64)}
    \STATE Cast $\boldsymbol{\theta}_{N_{\text{Adam}}} \rightarrow \boldsymbol{\theta}^{(64)}_{0}$, set $m \leftarrow 0$
    \WHILE {$m < M_{\max}$ \AND $\|\nabla_{\boldsymbol{\theta}}\mathcal{L}_{\text{total}}(\boldsymbol{\theta}_{m}^{(64)}; \mathcal{T}_{\text{L-BFGS}})\| \geq \tau$}
        \STATE \quad Update $\boldsymbol{\theta}^{(64)}$ using L-BFGS step:
        \STATE \qquad$\boldsymbol{\theta}^{(64)}_{m+1} \gets \text{L-BFGS}(\boldsymbol{\theta}^{(64)}_m,\, \mathcal{L}_{\text{total}}(\;\cdot \;; \mathcal{T}_{\text{L-BFGS}})),$
        \STATE \qquad $m \gets m+1$
    \ENDWHILE
    \vspace{1mm}
    \STATE \textbf{return} $\boldsymbol{\theta}_{\text{final}} \leftarrow \boldsymbol{\theta}_m^{(64)}$ 
\end{algorithmic}
\end{algorithm}
In the first optimization stage, the Adam optimizer~\cite{Kingma.2014} minimizes the composite loss function $\mathcal{L}_{\text{total}}(\boldsymbol{\theta})$ defined in Eq.~\eqref{eq:total_loss}. Adam employs randomized resampling of collocation points at a constant learning rate of $\eta = 2 \times 10^{-5}$ to ensure stable convergence. At every optimization step, a new set of $N_f$ spatial points $\mathcal{T}^{(n)}$ is randomly sampled from the normalized domain $(\bar{x}, \bar{y}) \in [-1, 1]^2$ to evaluate the PDE residuals. The stochastic sampling prevents the network from adapting to a fixed spatial pattern. To concentrate collocation points in regions of high residual error while maintaining a bounded point budget for L-BFGS optimization, a hybrid RAR strategy is employed. RAR evaluates the squared PDE residual magnitude across a dense, fixed candidate set $\mathcal{S}_{\text{cand}} \subset [-1, 1]^2$ using the parameters learned during Phase 1. The points with the largest residual errors are identified and merged with a subset of points drawn uniformly from the domain. This prioritizes regions where physical constraints are most violated. The collocation points identified in Phase 2 serve as the fixed evaluation set for Phase 3 fine-tuning. Bounding the collocation point budget is what makes the subsequent full-batch quasi-Newton step tractable, since standard L-BFGS requires the residual to be evaluated on a fixed set at every iteration. The final optimization phase employs the L-BFGS quasi-Newton algorithm for fine-tuning. L-BFGS approximates the inverse Hessian of the loss function, enabling faster convergence to high precision compared to first-order methods, at increased computational costs. This phase is essential since spectral bias is a dynamical property of the training process and not solely a limitation of the model representation. Resolving high-frequency content requires curvature-aware optimization, where first-order methods tend to stall on the high-frequency components of the solution~\cite{Khodakarami.2026}. The quasi-Newton update in L-BFGS relies on accurate gradients. The MR residuals contain second-order operators whose spectral contributions span several orders of magnitude. In single precision arithmetic (FP32) the resulting gradient noise can satisfy the L-BFGS convergence criterion prematurely. Therefore, all model parameters and arithmetic operations during the L-BFGS phase are cast to double precision (FP64)~\cite{Kiyani.2025, Xu.2025b}.

\subsection{Numerical Dataset}
\label{subsec:Numerical_Dataset}
The numerical dataset is generated through three-dimensional FE simulations in Abaqus/Explicit~\cite{Abaqus.2024}. The structure under consideration is a homogeneous, isotropic, linearly elastic aluminum plate of dimensions $(L \times L \times h) = 200 \times 200 \times 3\,\mathrm{mm}$ (EN AW-5083) with $E = 70\,\mathrm{GPa}$, $\nu = 0.33$ and $\rho = 2660\,\mathrm{kg/m}^3$. The plate is excited with a five-cycle Hann-windowed burst at a driving frequency of $100\,\mathrm{kHz}$. The dominant response frequency, determined from the measured wavefield spectrum is $f_0 = 96.4\,\mathrm{kHz}$. A mesh independence study confirmed convergence at 40 elements per shortest wavelength of the $A_0$ mode at the excitation frequency. Therefore, the plate is discretized with reduced-integration, eight-node brick elements (C3D8R) at this resolution, yielding a total of $980{,}000$ elements. All four plate edges are simulated traction-free. Parametric studies varying the Rayleigh damping coefficients (mass-proportional and stiffness-proportional) showed that an undamped model reproduces the experimental wavefield amplitudes and phases most accurately. The explicit time-domain simulation runs for $T = 300\,\mu\mathrm{s}$ utilizing automatic time increments with $\Delta t \approx 0.05\,\mathrm{\mu s}$. The excitation is applied as a concentrated force in the $z$-direction at the actuator position $(x,\,y) = (100,\,100)\,\mathrm{mm}$. The time-domain force history is taken from the experiment: the actuator displacement trace measured at the excitation location is spatially averaged over a $5\,\mathrm{mm}$-radius ring centered at $(100,\,100)\,\mathrm{mm}$, bandpass filtered between $30$ and $150\,\mathrm{kHz}$ and normalized before being loaded into the FE model. The actuator is driven with a five-cycle Hann-windowed sinusoidal burst at a center frequency of $100\,\mathrm{kHz}$. The complex amplitude fields $(\hat{\bar{u}},\,\hat{\bar{v}},\,\hat{\bar{w}})$ are extracted at the dominant response frequency $f_0 = 89.9 \,\mathrm{kHz}$ from the steady-state portion of the simulation. The resulting dataset provides the complex-valued transverse displacement $\hat{w}(\boldsymbol{x},\,f_0)$ on the FE mesh nodes over the full $200 \times 200\,\mathrm{mm}$ plate domain. From this field the normalized transverse displacement $\hat{\bar{w}}$ is obtained through the amplitude scaling $W_s$. This reference wavefield serves as the ground-truth baseline for training the PI-SIREN surrogate for the numerical validation of the damage identification pipeline and for comparison against the experimental dataset.

\subsection{Experimental Dataset}
\label{subsec:Experimental_Dataset}
\begin{figure*}[t]
\centering
\includegraphics[width=0.90\textwidth]{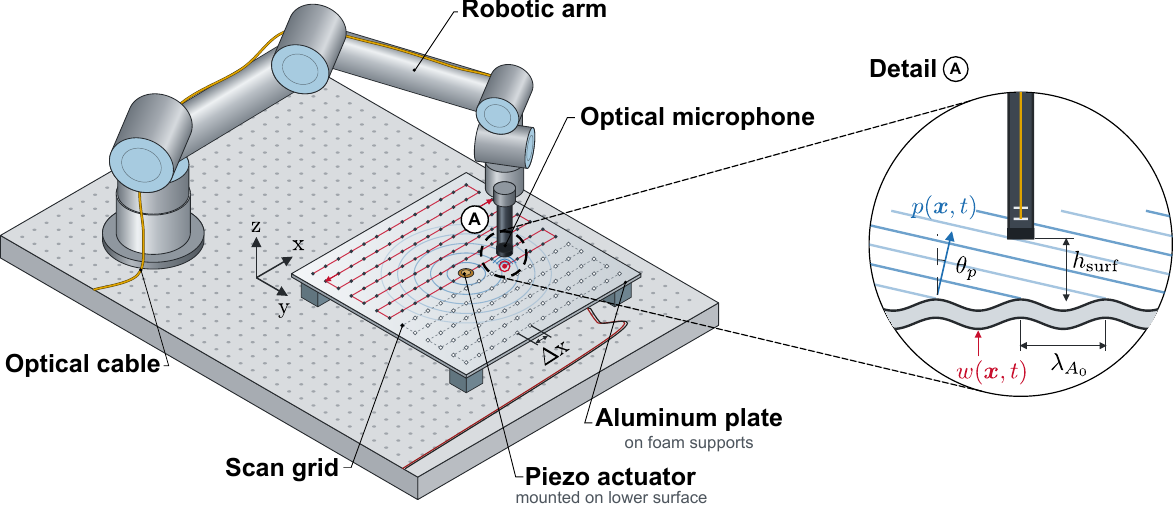}
\caption{Schematic representation of the experimental robotic measurement setup.}
\label{fig:Robtic_Measurement_Scetch}
\end{figure*}

The experimental campaign is performed on EN AW-5083 aluminum plates matching the numerical specimen material and geometry. Figure~\ref{fig:Robtic_Measurement_Scetch} illustrates the technical measurement setup used for this study. 
Excitation is achieved using a DuraAct piezoelectric transducer (cylindrical, $10\,\mathrm{mm}$ diameter, $0.25\,\mathrm{mm}$ thickness) vacuum-mounted at the plate center $(100, 100)\,\mathrm{mm}$. The actuator is driven through a dedicated power amplifier, delivering a $100\,\mathrm{kHz}$, five-cycle Hann-windowed sinusoidal burst. Surface vibrations are captured non-contactly using a Xarion ETA450 T-Shape optical microphone mounted on a UR-10 robotic manipulator. The microphone measures the acoustic pressure $p(\boldsymbol{x}, t)$ in the near field above the plate ($h_{\text{surf}}\approx 5\,\mathrm{mm}$), which is proportional to the normal surface velocity. The transverse displacement $w(\boldsymbol{x}, t)$ is obtained by time-integrating the pressure trace and dividing by the acoustic impedance of air $Z = \rho_0 c_0$. The incident angle between the microphone and plate surface is set to $90^{\circ}$. According to Snell's law, however, the plate radiates a leaky plane wave into the surrounding air at an angle of $\theta_p = \arcsin(c_{\text{air}} / c_p)$ from the surface normal~\cite{Rose.2014}. With $c_{\text{air}} = 343\,\mathrm{m/s}$ as the speed of sound at room temperature and $c_p\approx 1529\,\mathrm{m/s}$ as the phase velocity of the $A_0$-mode, obtained from dispersion relations, $\theta_p$ is calculated to $13^\circ$ and $\cos(\theta_p)$ is used as amplitude correction factor. The microphone output is recorded via a PicoScope 4842A oscilloscope. The raw measurement traces undergo the following pre-processing before frequency-domain analysis. Since the measurements extend beyond the plate edges, the spatial scan grid is co-registered with the plate geometry and cropped to the $200 \times 200\,\mathrm{mm}$ physical domain. The global time offset is estimated by averaging the traces at the grid points closest to the actuator and locating the onset of the first-arriving wave. Then, all traces are shifted so that $t=0$ coincides with the start of the excitation signal. Each trace undergoes bandpass filtering with a fourth-order Butterworth filter (passband $30$--$150\,\mathrm{kHz}$) and truncation to $300\,\mu\mathrm{s}$ to match the numerical simulation duration. The pre-processed dataset comprises $N_d$ time-domain displacement traces recorded on a uniform spatial grid over the $200 \times 200\,\mathrm{mm}$ plate surface. From these traces the complex amplitude field $\hat{w}(\boldsymbol{x}, f_0)$ of the harmonic component at $f_0 = 96.4\,\mathrm{kHz}$ is obtained from the discrete Fourier transform (DFT), providing the experimental reference for training and validation. 

\subsection{Damage Identification Methodology}
\label{subsec:damage_identification}
\begin{figure*}[t]
\centering
\includegraphics[width=0.95\textwidth]{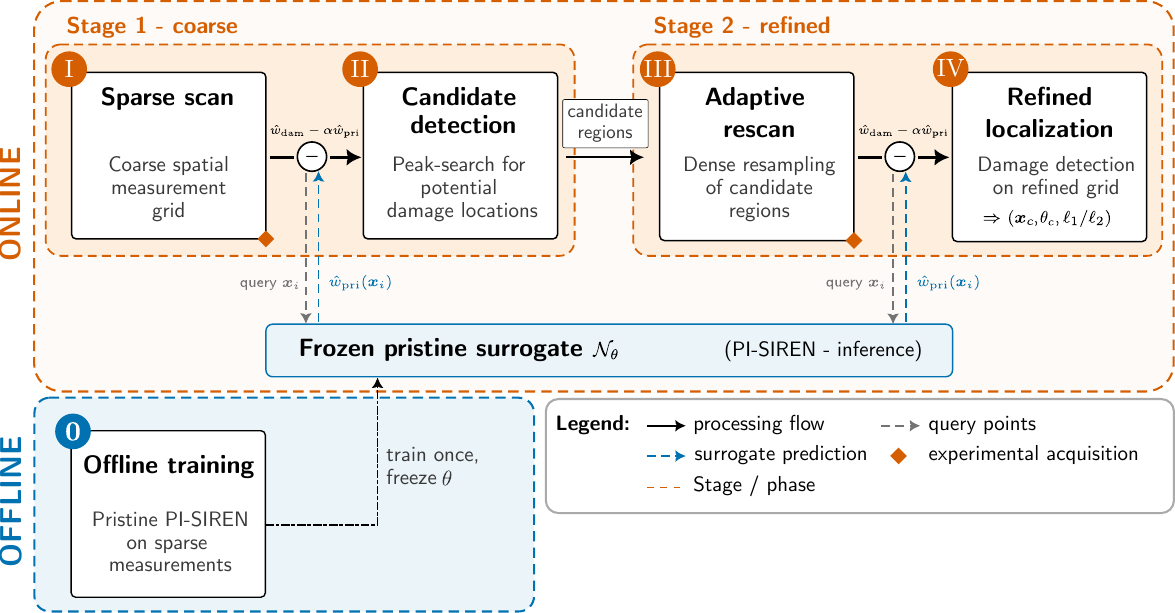}
\caption{Sequential damage identification pipeline on experimental data: The pristine baseline model is trained once offline and reused as pre-trained model in every online inspection for inference. Online phases~I--IV: coarse field acquisition, baseline subtraction and candidate detection, adaptive resampling and refined damage localization. }
\label{fig:Damage_identification_pipeline}
\end{figure*}

The trained pristine PI-SIREN provides a continuous, physics-consistent representation of the undamaged wavefield that can be evaluated at arbitrary spatial locations without retraining. Damage is detected by subtracting this baseline from a new measurement. The full procedure is illustrated in Figure~\ref{fig:Damage_identification_pipeline} and comprises an offline training phase followed by a sequential online pipeline: a coarse-grid evaluation and an adaptive refinement evaluation of the surrogate.
Let $\hat{w}_{\text{dam}}(\boldsymbol{x})$ denote the complex transverse displacement amplitude of the damaged plate at $\boldsymbol{x}=(x,y)$, obtained from measurement or FE simulations and let $\hat{w}_{\text{pri}}(\boldsymbol{x})$ be the corresponding pristine PI-SIREN prediction. Minor global amplitude and phase discrepancies are removed by fitting the pristine field to the damaged field through a single complex scalar $\alpha$:
\begin{multline}
    \alpha
    = \operatorname*{arg\,min}_{\alpha \in \mathbb{C}}
      \bigl\| \hat{w}_{\text{dam}} - \alpha\,\hat{w}_{\text{pri}} \bigr\|_{2}^{2} \\
       =  \frac{\displaystyle\sum_{\boldsymbol{x} \in \Omega_v}
      \hat{w}_{\text{dam}}(\boldsymbol{x})\,
      \overline{\hat{w}_{\text{pri}}(\boldsymbol{x})}}
      {\displaystyle\sum_{\boldsymbol{x} \in \Omega_v}
      \bigl|\hat{w}_{\text{pri}}(\boldsymbol{x})\bigr|^{2}},
    \label{eq:complex_alignment}
\end{multline}
where $\Omega_v$ is the set of valid measurement points and $\overline{(\cdot)}$ denotes complex conjugation. The scattered field is the residual
\begin{equation}
    \hat{w}_{\text{scat}}(\boldsymbol{x}_i)
    = \hat{w}_{\text{dam}}(\boldsymbol{x}_i)
    - \alpha\,\hat{w}_{\text{pri}}(\boldsymbol{x}_i),
    \label{eq:scattered_field}
\end{equation}
with $i = 1, \dots, N_s$, evaluated at $N_s$ measurement points before interpolation. This subtraction suppresses the dominant incident wavefield, retaining the weaker scattering signature of the defect. 

The first evaluation, phase~(I) of Figure~\ref{fig:Damage_identification_pipeline}, operates on a sparse uniform measurement grid. The scattered values are interpolated onto a Cartesian grid via Clough--Tocher~\cite{Mann.1999} cubic interpolation, $\widetilde{w}_{\text{scat}} = \mathcal{I}[\hat{w}_{\text{scat}}]$ and smoothed with an isotropic Gaussian kernel of width $\sigma$ to yield $W_\sigma = G_\sigma * \widetilde{w}_{\text{scat}}$. Three complementary indicators are then evaluated. The first two act on the scattered field, while the third compares damaged and pristine fields locally:
\begin{align}
    \mathcal{D}_{\text{mag}}(\boldsymbol{x})
      &= \bigl| W_{\sigma}(\boldsymbol{x}) \bigr|, \\
    \mathcal{D}_{\mathcal{H}}(\boldsymbol{x})
      &= \bigl| \mathrm{Im}\{\, \overline{W_{\sigma}}\,
        \nabla^{2} W_{\sigma} \,\} \bigr|, \\
    \mathcal{D}_{\text{inc}}(\boldsymbol{x})
      &= 1 - \frac{\bigl| \bigl\langle \hat{w}_{\text{dam}}\,
        \overline{\hat{w}_{\text{pri}}} \bigr\rangle_{B_R}(\boldsymbol{x}) \bigr|}
        {\bigl\langle \bigl| \hat{w}_{\text{dam}} \bigr| \bigr\rangle_{B_R}(\boldsymbol{x})
         \; \bigl\langle \bigl| \hat{w}_{\text{pri}} \bigr| \bigr\rangle_{B_R}(\boldsymbol{x})},
    \label{eq:indicators}
\end{align}
where $\langle\,\cdot\,\rangle_{B_R}$ denotes a box average over a square window of half-width $R$ and $\hat{w}_{\text{pri}}$ is the aligned pristine field $\alpha$-scaled according to Eq.~\eqref{eq:complex_alignment}. The three indicators are complementary: $\mathcal{D}_{\text{mag}}$ is robust to noise but poorly localized, $\mathcal{D}_{\mathcal{H}}$ is sharply localized but derivative-based and noise-sensitive and $\mathcal{D}_{\text{inc}}$ depends only on local phase coherence and is insensitive to amplitude calibration. The Helmholtz-type indicator is proportional to the divergence of the time-averaged flexural intensity flux, $\nabla\cdot\mathbf{I} \propto \mathrm{Im}\{\overline{\hat{w}}\,\nabla^2\hat{w}\}$, which for the MR plate reduces to $-\mathrm{Im}\{\overline{\hat{w}}\,(\nabla\cdot\boldsymbol{\psi})\}$. Fusing the three indicators reduces the risk that a defect is missed due to measurement conditions that degrade a single indicator.

To suppress smooth background residuals, each indicator is normalized via a rectified $z$-score over a sliding box window of half-width $R$. The rectified $z$-score
\begin{equation}
    Z(\boldsymbol{x}) = \max\!\left(
      \frac{\mathcal{D}(\boldsymbol{x}) - \mu_R(\boldsymbol{x})}
           {\varsigma_R(\boldsymbol{x})},\; 0 \right) \chi_v(\boldsymbol{x})
    \label{eq:zscore}
\end{equation}
retains only positive deviations from the local background, where $\chi_v$ masks the valid measurement domain. The three $z$-scored indicators are rescaled to $[0,1]$ and combined into a single fused detection score via an equal-weighted mean. In phase (II), candidate damage regions are identified as local maxima of this fused score that exceed a quantile threshold. A non-maximum suppression step enforces a minimum spatial separation between candidates to avoid clustered duplicates. The remaining candidates are ranked by score and define rectangular regions of interest. These regions are adaptively rescanned on a finer grid in phase (III), while the original coarse points are retained. This concentrates the measurement effort where anomalies were initially detected. 

In the second evaluation, phase~(III) and~(IV), the pre-trained PI-SIREN is reevaluated at all measurement points (coarse and resampled) and the scattered field is recomputed. The fused score serves exclusively for candidate detection in the first pass. For the final localization, only $\mathcal{D}_{\mathcal{H}}$ is evaluated, since at the refined sampling density its derivative-based sharpness resolves the defect centroid more precisely than the fused score. The dominant response region is isolated by thresholding $\mathcal{D}_{\mathcal{H}}$, yielding the region $\Omega_b$. The damage geometry within $\Omega_b$ is characterized by intensity-weighted image moments. With $\mathcal{S}(\boldsymbol{x})$ denoting the normalized indicator and $m_0 = \sum_{\boldsymbol{x}\in\Omega_b}\mathcal{S}(\boldsymbol{x})$ its integrated mass, the centroid and covariance matrix are
\begin{align}
    \boldsymbol{x}_c &= \frac{1}{m_0}
      \sum_{\boldsymbol{x}\in\Omega_b} \mathcal{S}(\boldsymbol{x})\,
      \boldsymbol{x}, \label{eq:center} \\
    \boldsymbol{\Sigma} &= \frac{1}{m_0}
      \sum_{\boldsymbol{x}\in\Omega_b} \mathcal{S}(\boldsymbol{x})\,
      (\boldsymbol{x}-\boldsymbol{x}_c)(\boldsymbol{x}-\boldsymbol{x}_c)^{\!\top}.
    \label{eq:moments}
\end{align}
The centroid $\boldsymbol{x}_c$ gives the damage location. The eigenvectors of $\boldsymbol{\Sigma}$ yield the principal orientation $\theta_c$ and the eigenvalues $\varrho_1 \ge \varrho_2$ give the characteristic extents $\ell_k = 2\sqrt{3}\,\sqrt{\varrho_k}$. The pipeline outputs $\boldsymbol{x}_c$, $\theta_c$ and the aspect ratio $\ell_1/\ell_2$ of the damage.

\section{Results}
\label{sec:results}
This section evaluates the proposed PI-SIREN against the numerical and experimental datasets. The proposed optimization strategy (Adam, RAR, L-BFGS) for pristine wavefield reconstruction using FE simulation data is established in the subsection on numerical wavefield reconstruction. Next, the minimum spatial sensor density required for reliable damage localization is examined using the same FE dataset. The contribution of the network architecture is then isolated by applying the resulting configuration to conventional and Fourier-feature MLP-$\tanh$ baselines. The resulting sensor density and architecture are used in the experimental campaign, where the pristine reconstruction is validated against measured wavefields and the two-stage damage identification pipeline is applied to experimental data containing an introduced defect.

\subsection{Numerical Wavefield Reconstruction}
\label{sec:res_recon}
In this section, the proposed optimization strategy (Algorithm~\ref{alg:FreqDomain_PINN}) is analyzed and evaluated in terms of its effectiveness as part of an ablation study. The strategy is evaluated using four training configurations labeled M1--M4 in
\begin{table*}[!t]
\centering
\caption{Reconstruction ablation study on the $25\times25$ grid. Relative test error $L_2(\hat{\bar{w}})$ on a fixed dense test set and the final PDE residual $\mathcal{L}_{\text{PDE}}$.}
\label{tab:recon_ablation}
\small
\begin{tabular}{@{}clcc@{}}
\toprule
Tag & Configuration & $L_2(\hat{\bar{w}})$ & $\mathcal{L}_{\text{PDE}}$ \\
\midrule
M1 & Data-only ($\lambda_{\text{PDE}}=\lambda_{\text{BC}}=0$) & 0.4932 & -- \\
M2 & PINN, Adam-only (70{,}000 iter.) & 0.2881 & $2.7\times10^{-4}$ \\
M3 & PINN, Adam + L-BFGS & 0.1351 & $5.2\times10^{-6}$ \\
M4 & PINN, Adam + RAR + L-BFGS & 0.1037 & $5.6\times10^{-6}$ \\
\bottomrule
\end{tabular}
\end{table*}
Table~\ref{tab:recon_ablation}, which differ by a single component. This allows the attribution of any change in accuracy to the additional component. M1 removes the physics ($\lambda_{\text{PDE}}=\lambda_{\text{BC}}=0$), reducing the model to a data-driven interpolant. M2 uses the full physics-informed loss but optimizes with Adam only, isolating the quasi-Newton (L-BFGS) contribution. M3 adds the L-BFGS phase, but the L-BFGS collocation points are drawn uniformly at random to isolate the RAR contribution. M4 corresponds to the proposed pipeline: Adam, RAR and L-BFGS in FP64. All four models share the same data, architecture, loss weights and number of collocation points.
\begin{table}[!b]
\centering
\caption{Training configuration held constant across all numerical and experimental tests unless stated otherwise.}
\label{tab:hparams}
\small
\begin{tabular}{@{}ll@{}}
\toprule
Hyperparameter & Value \\
\midrule
Hidden layers / neurons & $5 \times 256$ \\
Base frequency $\omega_0$ (all layers) & 30 \\
Data supervision & transverse $\hat{\bar{w}}$ only \\
Loss weights $(\lambda_{\text{PDE}},\, \lambda_{\text{BC}},\, \lambda_{\text{data}})$ & $(1.0,\,\! 10.0,\,\! 10.0)$ \\
Adam learning rate $\eta$ & $2 \times 10^{-5}$ \\
Adam iterations (FP32) & 20{,}000 \\
RAR: candidate set size $N_{\text{cand}}$ & 20{,}000 \\
RAR: hard / random split & 50\,\% / 50\,\% \\
L-BFGS max iterations (FP64) $M_{\max}$ & 50{,}000 \\
\bottomrule
\end{tabular}
\end{table}
The model is trained on a $25\times25$ spatial grid ($\Delta x \approx 8\,\mathrm{mm}$), which provides approximately two measurement points per $A_0$-mode wavelength at $f_0$, the Nyquist-limit data density. The sufficiency of this grid for the downstream localization task is established in the following density study. The training configuration is held constant across all variants (Table~\ref{tab:hparams}). All variants are evaluated on the same fixed test dataset. 

Table~\ref{tab:recon_ablation} shows the ablation results. The full optimization pipeline (M4) reaches a test $L_2(\hat{\bar{w}})$-error of $0.104$, representing reductions of $79\,\%$, $64\,\%$ and $23\,\%$ relative to the Data-only (M1, $0.493$), Adam-only (M2, $0.288$) and no-RAR (M3, $0.135$) runs, respectively. The M1 result indicates that the PDE residual provides dominant regularization. With 625 measurement points, the Data-only model is insufficient to constrain the ultrasonic wavefield at $f_0$ without physical regularization. The final PDE residual of M2 remains nearly two orders of magnitude above the M3--M4 values ($\sim 5\times10^{-6}$) due to the L-BFGS phase. The step from M2 to M3 halves the $L_2$ error, identifying the L-BFGS phase as the main source of increased accuracy. The FP64 second-order refinement reaches minima that the first-order Adam does not approach within the same number of iterations. The final step, from M3 to M4, attributes a $23\,\%$ reduction to RAR. One plausible explanation is that L-BFGS collocation concentrates on high-residual regions, providing more informative curvature samples than uniform sampling, which allocates points to already-converged, low-residual regions. 

\begin{figure}[htbp]
\centering
\includegraphics[width=1\linewidth]{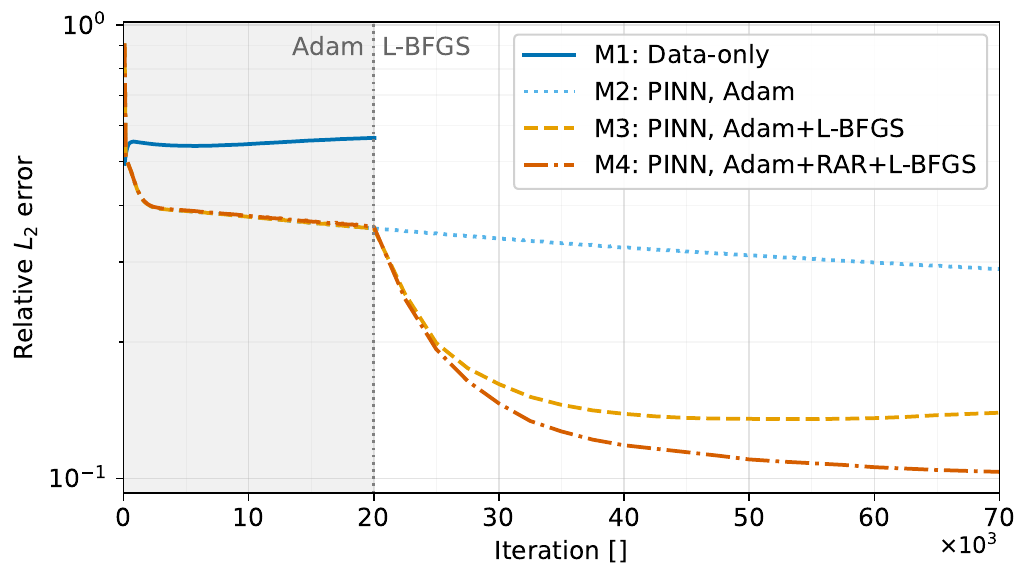}
\caption{Test $L_2$ versus training iteration for all four ablation variants on the $25^2$ grid.}
\label{fig:conv}
\end{figure}

\begin{figure*}[htbp]
\centering
\includegraphics[width=1\linewidth]{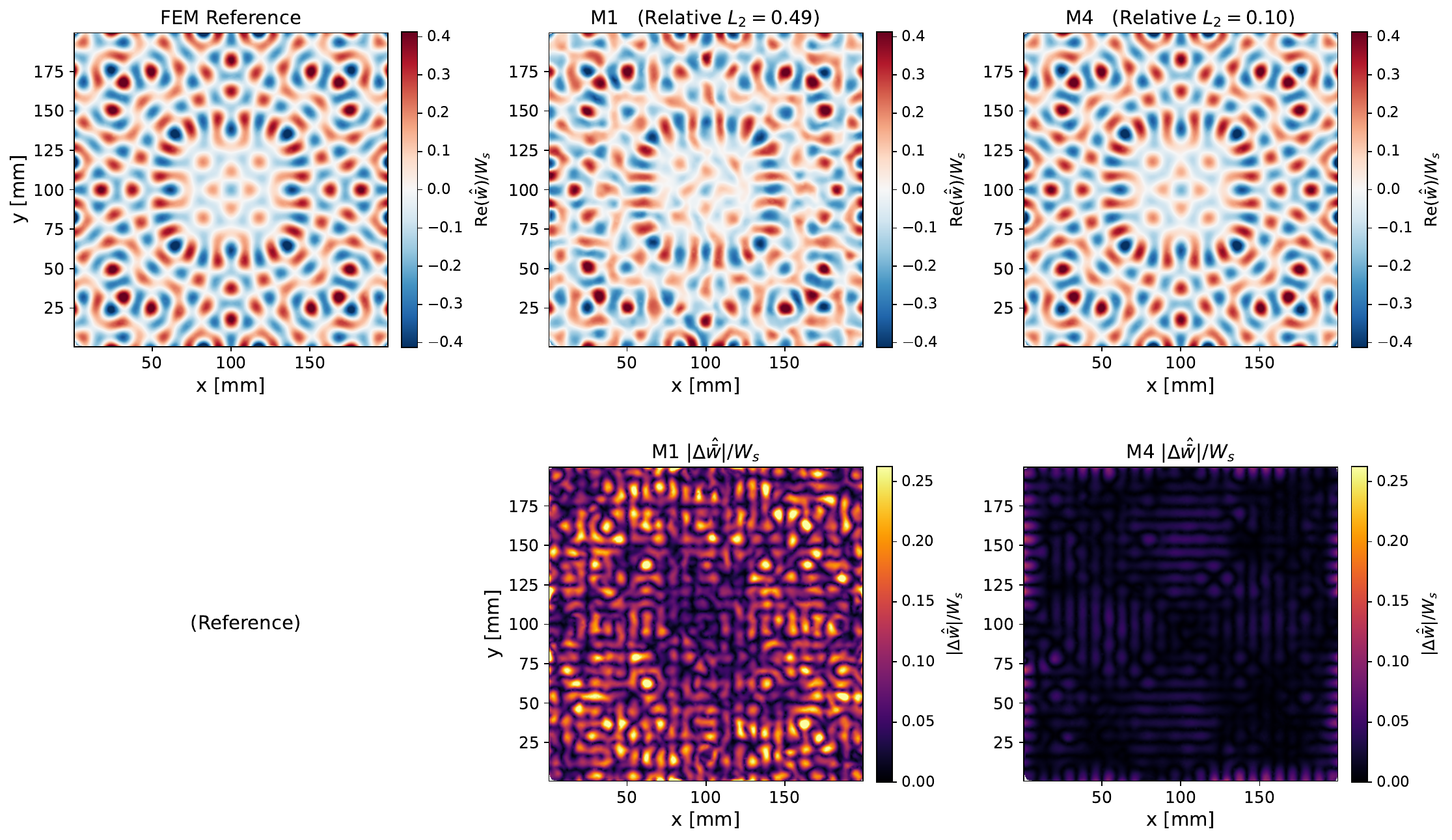}
\caption{Pristine wavefield reconstruction on the $25^2$ grid: FEM reference (left), M1 with $L_2 = 0.49$ (center) and M4 PI-SIREN prediction with $L_2 = 0.10$ (right). Top row: normalized transverse displacement $\mathrm{Re}(\hat{\bar{w}})/W_s$, bottom row: pointwise error $|\Delta \hat{\bar{w}}|/W_s$.}
\label{fig:recon_field}
\end{figure*}

Figure~\ref{fig:conv} shows the convergence trajectories of the relative $L_2$ error versus the number of iterations. During the Adam phase, M3 and M4 decrease at a comparable rate and plateau near $L_2\approx0.1$--$0.15$. The Data-only model M1 plateaus near $L_2 \approx 0.3$--$0.5$. M2 tracks the physics-informed variants through the Adam phase. The transition to L-BFGS in M3 and M4 produces a significant drop in $L_2$-error driven by quasi Newton optimization with FP64 precision.  
Figure~\ref{fig:recon_field} compares the reconstructed field against the FE reference for model M4 (full optimization strategy) and M1 (Data-only). M4 is able to reconstruct the emitted wavefield with low error. The real part of the transverse displacement and pointwise error are shown on a common, normalized scale by the amplitude scaling factor $W_s$. The pointwise normalized error, defined as $|\Delta w|/W_s$, remains below $0.1$ over most of the domain. Minor locally elevated errors occur near the plate edges, where the free-boundary condition is most sensitive to the collocation density. 
The full optimization strategy (M4: Adam, RAR and L-BFGS) is adopted and held constant for all subsequent results. Only the training grid changes in the following sensor density study, which establishes the data requirement of the surrogate. All hyperparameters and loss weights are the same as those specified in Table~\ref{tab:hparams}.

\subsection{Sensor Density Requirement}
\label{sec:res_density}
\begin{figure}[htbp]
\centering
\includegraphics[width=1\linewidth]{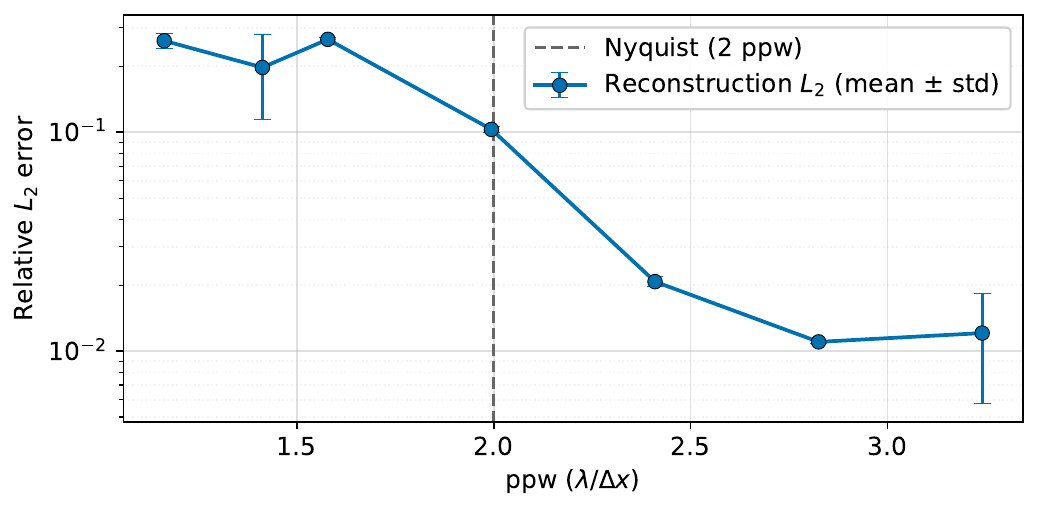}
\caption{Reconstruction $L_2(\hat{\bar{w}})$ versus sensor density in ppw (mean~$\pm$~std).}
\label{fig:l2_ppw}
\end{figure}

The architecture, loss formulation and optimization strategy defined in the Methodology and validated in the preceding subsection determine how the PI-SIREN is trained, but not how many measurement points the training requires for sufficient damage identification. In a practical ultrasonic inspection task, the scan time scales linearly with the number of sampled points, making the spatial sensor density one of the dominant constraints. This section identifies the coarsest uniform training grid for which the surrogate can reliably localize damage using FE data. This differs from the question of how small the global $L_2$ error can be made for purely reconstruction purposes: a wavefield can be reconstructed with a moderate global error yet still preserve the local scattering signature of a defect. 
The PI-SIREN is retrained on a series of uniform grids from $15\times15$ to $40\times40$ points over the $200\times200\,\mathrm{mm}$ plate domain, using the optimization strategy and hyperparameters of M4 fixed in Table~\ref{tab:hparams}. The grid spacing is $\Delta x = L/(N{-}1)$ and the sampling density is expressed as points per wavelength ($\mathrm{ppw} = \lambda/\Delta x$), where $\lambda \approx 16.6\,\mathrm{mm}$ is the $A_0$-mode wavelength extracted from the FEM reference field at $f_0 = 89.9\,\mathrm{kHz}$. The Nyquist limit corresponds to 2~ppw, approximated by the $25\times25$ grid. Two metrics are tracked simultaneously: (i)~the relative $L_2$ error of the transverse displacement on a fixed test dataset, measuring global reconstruction accuracy and (ii)~the localization error, measuring damage identification accuracy. The damaged reference fields are generated by removing all FEM elements within a line region of $8\,\mathrm{mm}$ at four ($x, y$) positions: $(130,50)$, $(150,50)$, $(170,100)$ and $(30,30)\,\mathrm{mm}$, to cover different scattering geometries relative to the centrally mounted actuator. Three random initialization seeds are used per grid to account for optimization variability. A localization is counted as successful when the localization error is below $2\,\mathrm{mm}$.
\begin{table}[htbp]
\centering
\caption{Sensor density study: ppw and pristine-field reconstruction error $L_2(\hat{\bar{w}})$ (mean~$\pm$~std over three seeds) for each uniform training grid.}
\label{tab:ppw_study}
\small
\begin{tabular}{@{}lcc@{}}
\toprule
Grid & ppw & $L_2(\hat{\bar{w}})$ \\
\midrule
$15\times15$ & 1.16 & $0.262 \pm 0.0208$ \\
$18\times18$ & 1.41 & $0.198 \pm 0.0834$ \\
$20\times20$ & 1.58 & $0.265 \pm 0.0062$ \\
$25\times25$ & 1.99 & $0.103 \pm 0.0032$ \\
$30\times30$ & 2.41 & $0.021 \pm 0.0011$ \\
$35\times35$ & 2.83 & $0.011 \pm 0.0001$ \\
$40\times40$ & 3.24 & $0.012 \pm 0.0063$ \\
\bottomrule
\end{tabular}
\end{table}
Table~\ref{tab:ppw_study} and Figure~\ref{fig:l2_ppw} report the reconstruction error as a function of sensor density. The $L_2(\hat{\bar{w}})$ error decreases from approximately $26\,\%$ at 1.2~ppw to about $10\,\%$ at the Nyquist limit ($25^2$, 2.0~ppw) and continues to fall past Nyquist to $\sim 2\,\%$ at 2.4~ppw ($30^2$) and $\sim 1\,\%$ at 2.8~ppw ($35^2$). 

\begin{figure}[htbp]
\centering
\includegraphics[width=0.95\linewidth]{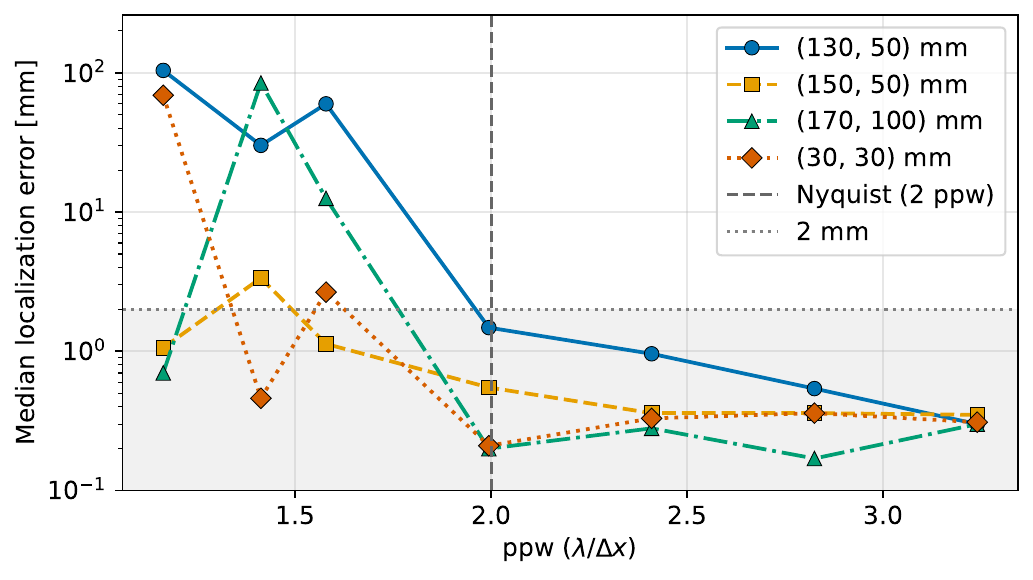}
\caption{Median localization error versus sensor density in ppw for four simulated damage positions with $2\,\mathrm{mm}$ acceptance threshold. Above the Nyquist limit (dashed line, 2~ppw) all positions converge to $< 1\,\mathrm{mm}$.}
\label{fig:positions}
\end{figure}

\begin{figure}[htbp]
\centering
\includegraphics[width=1\linewidth]{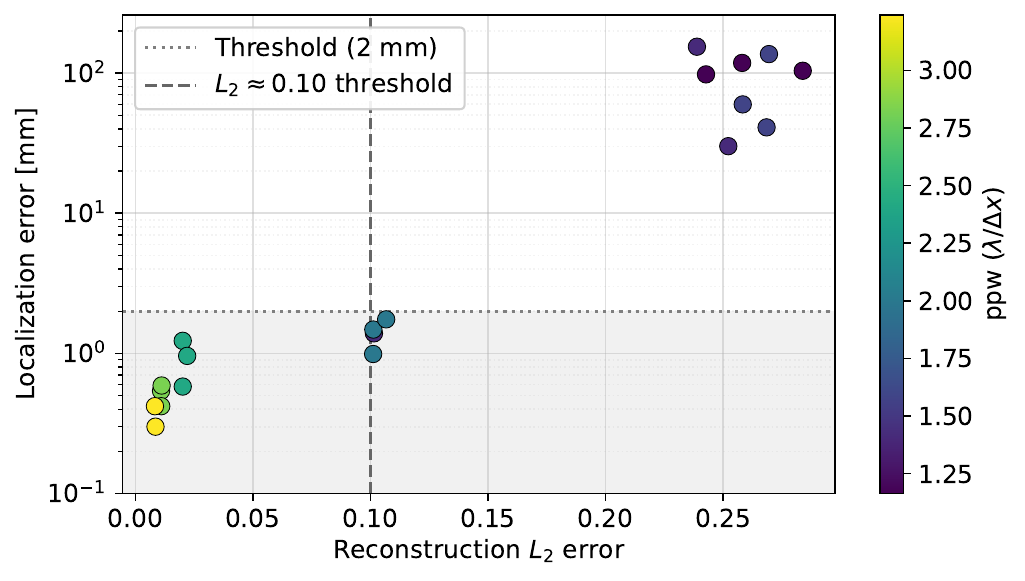}
\caption{Localization error versus reconstruction $L_2(\hat{\bar{w}})$ with damage at $(130,50)\,\mathrm{mm}$. Successful localizations occur predominantly for $L_2 \lesssim 0.10$.}
\label{fig:l2_loc}
\end{figure}

Beyond 2.8~ppw the error curve flattens: the $40^2$ grid (3.2~ppw) yields $L_2 \approx 1.2\,\%$, a slightly increased error in comparison to the $35^2$ result. The reconstruction error therefore saturates near 2.8~ppw and further densification yields no measurable gain in error reduction. Notably, this saturation point lies above the Nyquist density: the full-field reconstruction continues to improve for nearly one additional ppw beyond the density at which damage localization already becomes reliable (cf.~Figure~\ref{fig:positions}).
For each trained model on FE simulation data, the localization error is determined from the final single-pass evaluation of the damage identification pipeline (phase~(III) and~(IV) in Figure~\ref{fig:Damage_identification_pipeline}). The pristine surrogate prediction is aligned to the damaged field via the complex scalar $\alpha$ (Eq.~\eqref{eq:complex_alignment}), the scattered field is obtained by subtraction and the damage location is extracted from the peak of the Gaussian-smoothed intensity-divergence indicator $\mathcal{D} = \bigl|\mathrm{Im}\{\overline{W_{\sigma}}\,\nabla^{2} W_{\sigma}\}\bigr|$. The resulting damage localization error as a function of sensor density (ppw) is shown in Figure~\ref{fig:positions}. At and above $25^2$ points (2.0~ppw), the median error for all four positions remains below the $2\,\mathrm{mm}$ acceptance threshold across all seeds. Below this density, the median error becomes position-dependent and non-monotonic in grid size. For instance, the defect at $(170,100)\,\mathrm{mm}$ stays below $2\,\mathrm{mm}$ at $15^2$ (1.16~ppw) but exceeds it at $18^2$ and $20^2$, while the defect at $(30,30)\,\mathrm{mm}$ shows the reverse trend (Figure~\ref{fig:positions}). The non-monotonic success pattern is physically consistent with spatial aliasing. A sub-Nyquist grid cannot uniquely represent the continuous wavefield: high spatial frequencies fold back into the baseband and the resulting aliased field depends on the relative phase between the sampling grid and the wavefront at each location. For a given defect, the aliased reconstruction may either preserve or suppress the local scattering contrast, depending on this phase relationship. Below the Nyquist limit, the localization outcome is governed by the geometric relationship between the defect position and the sampling grid rather than by the sampling density itself.
A clearer picture emerges when reconstruction and localization accuracy are plotted against each other. Figure~\ref{fig:l2_loc} shows the localization error as a function of the reconstruction $L_2(\hat{\bar{w}})$ for the defect at $(130,50)\,\mathrm{mm}$, with color encoding the sensor density. A sharp bimodal separation is observed: runs with $L_2 \lesssim 0.10$ yield localization errors below the 2~mm acceptance threshold, whereas runs with $L_2 \gtrsim 0.24$ miss by more than $25\,\mathrm{mm}$. The vertical line at $L_2 = 0.10$ separates the two clusters.
The few sub-Nyquist successes correspond to model runs in which the aliasing pattern happened to preserve the scattered-field contrast, yielding a per-run reconstruction $L_2$ error at the lower end of the inter-seed spread. Conversely, the sub-Nyquist failures are runs in which aliasing degraded the reconstruction accuracy, with an $L_2$ error above the threshold. Sensor density is thus a measure of reconstruction accuracy. It determines whether the reconstruction error is low enough for the localization pipeline to extract the defect signature. The contrast between the two saturation points is the central finding of this section: localization becomes reliable at the Nyquist density (2.0~ppw, $25^2$), while the full-field reconstruction continues to improve until approximately 2.8~ppw ($35^2$). The damage-identification task therefore requires less reconstruction accuracy than the global $L_2$ metric suggests, because the pipeline extracts a local scattering signature rather than reproducing the entire wavefield.

For the experimental demonstration, the nominal sensor density is set at the Nyquist limit (625 points, $25^2$, 2.0~ppw), with spatial robustness ensured through stratified subsampling of a coarse scan rather than a rigid grid. Although the $L_2$--ppw threshold varies with defect size, excitation frequency and plate geometry, the diagnostic from numerical data is transferable to experiments. 

\subsection{Architectural Comparison}
\label{sec:res_arch}
\begin{table*}[!tbp]
\centering
\caption{Architecture comparison on the $25\times25$ grid. $L_2$ values are mean~$\pm$~std over three seeds. The localization error is the median over the 12 runs (three seeds $\times$ four defect positions).}
\label{tab:architecture}
\small
\begin{adjustbox}{max width=\linewidth}
\begin{tabular}{@{}llllcccc@{}}
\toprule
Tag & Model & Activation & Input, $[\mathrm{Layers} \times \mathrm{Neurons}]$ & $N_{\text{params}}$ & $\eta$ & $L_2(\hat{\bar{w}})$ & Loc. $\|\Delta\boldsymbol{x}_c\|$ $[\mathrm{mm}]$ \\
\midrule
M4 & PI-SIREN & $\sin(\omega_0\xi)$ & coordinates, $[5\times256]$ & 265{,}478 & $2\times10^{-5}$ & $\mathbf{0.103\pm0.003}$ & $\mathbf{0.45}$ \\
M5 & MLP-tanh & $\tanh$ & coordinates, $[5\times256]$ & 265{,}478 & $10^{-3}$ & $0.490\pm0.039$ & $110.15$ \\
M6 & FF-MLP-tanh & $\tanh$ & Fourier Feature, $[5\times241]$ & 265{,}829 & $10^{-3}$ & $0.245\pm0.007$ & $58.85$ \\
\bottomrule
\end{tabular}
\end{adjustbox}
\end{table*}
The proposed PI-SIREN (M4) is compared against two MLP baselines with hyperbolic tangent activation functions (M5 and M6) of comparable size. Both models are trained using the same $25\times25$ grid with the same collocation and boundary budgets, the same loss weights and the same three-phase optimization strategy of Algorithm~\ref{alg:FreqDomain_PINN}. Three initialization seeds are used per model. Training parameters and results are summarized in Table~\ref{tab:architecture}. M5 is a fully connected network with $\tanh$ activation functions and no input embedding. M6 augments the input with a fixed random Fourier feature embedding~\cite{Tancik.2020} $\phi(\boldsymbol{\zeta})=[\sin(2\pi\mathbf{B}\boldsymbol{\zeta}), \cos(2\pi\mathbf{B}\boldsymbol{\zeta})]$, where the $64\times2$ matrix $\mathbf{B}$ is drawn in two bands of 32 frequencies each from $\mathcal{N}(0,\sigma_f^2)$, with $\sigma_f$ set to $2.0$ and $6.0$ cycles per computational unit. The embedding input $\boldsymbol{\zeta} = (\bar{x}, \bar{y})$ is the normalized coordinate vector. The embedding itself yields 128 input channels. The network width is reduced from 256 to 241 neurons per layer, so that the number of trainable parameters $N_{\text{params}}$ remains comparable to those of M4 and M5. Under the input normalization to $[-1,1]^2$, one computational unit corresponds to a physical length of $L/2$, so the plate spans two computational units. The $A_0$-mode wavelength of $\lambda\approx16.6\,\mathrm{mm}$ corresponds to $L/(2\lambda)\approx 6$ cycles per computational unit, which the $\sigma_f=6.0$ band targets at one standard deviation, corresponding to $k_{A_0}\approx0.38\,\mathrm{rad/mm}$, while the $\sigma_f=2.0$ band covers the wave envelope. Both comparison models M5 and M6 use the learning rate that yields the lowest test error for the Adam phase in an independent grid search, yielding $\eta=10^{-3}$. Additionally, M5 and M6 employ a 1{,}000-step warm-up during the Adam phase followed by a cosine decay to $10\,\%$ of the initial learning rate. 

As shown in Table~\ref{tab:architecture}, the PI-SIREN reduces the reconstruction error by $58\,\%$ relative to the Fourier-feature baseline (M6) and by $79\,\%$ relative to the plain MLP-$\tanh$ network (M5). The error of M5 matches that of the data-only PI-SIREN M1, which reaches $0.4932$ with the physics entirely removed (Table~\ref{tab:recon_ablation}). A $\tanh$ network equipped with the full MR residual thus performs on par with a SIREN trained without any physical constraint. M6 halves the error of M5, confirming that enriching the input spectrum is effective, yet its error still exceeds that of M4 by a factor of 2.4.

\begin{figure*}[!tbp]
\centering
\includegraphics[width=1\textwidth]{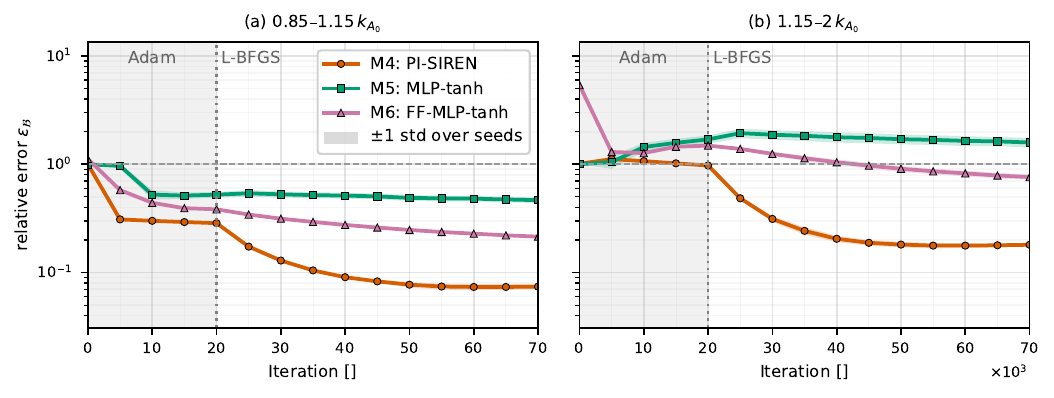}
\caption{Relative band error $\varepsilon_{\mathcal{B}}$ against training iteration: (a)~$A_0$ band, $0.85\,k_{A_0}\le|\mathbf{k}|\le1.15\,k_{A_0}$, and (b)~short-wavelength band above the $A_0$ band, $1.15\,k_{A_0}<|\mathbf{k}|\le2.0\,k_{A_0}$.}
\label{fig:spectral_dynamics}
\end{figure*}

The trained models are evaluated as pristine baselines on the same four defect positions as shown in Figure~\ref{fig:positions}. As in the sensor density study, the damage indicator is computed in a single pass on a dense grid of $10{,}000$ sampled points without adaptive refinement. All M4 inferences localize the defects within the $2\,\mathrm{mm}$ threshold with a median localization error of $0.45\,\mathrm{mm}$, whereas neither of the MLP baselines is able to localize a single defect. The failures of M5 are largely independent of the defect position, as a given seed returns the same predicted location for all four damage cases. The detected peak therefore responds to the systematic reconstruction error of the architecture rather than to a scattering signature that the residual field no longer contains. This is supported by the magnitude of the alignment factor in Eq.~\eqref{eq:complex_alignment}, which stays within $0.95$--$1.00$ for M4 and $1.00$--$1.05$ for M6 but drops to $0.83$--$0.93$ for M5. This indicates a systematic, spatially varying amplitude bias that the single scalar alignment cannot remove. Unlike the sub-Nyquist failures identified in the sensor density study, which arise from spatial aliasing of a field that still resolves the $A_0$ mode, these failures indicate a reconstruction that does not reproduce the $A_0$ mode with sufficient accuracy. To identify the spectral origin of this failure, the reconstruction error is evaluated in two wavenumber bands around $k_{A_0}=2\pi/\lambda_{A_0}$: the $A_0$ band at $0.85\,k_{A_0}\le|\mathbf{k}|\le1.15\,k_{A_0}$, which contains the propagating wave and the short-wavelength band at $1.15\,k_{A_0}<|\mathbf{k}|\le2.0\,k_{A_0}$, which contains the higher spatial frequencies. For a band $\mathcal{B}$ the relative error is defined as
\begin{equation}
    \varepsilon_{\mathcal{B}} = 
    \frac{\bigl\lVert\hat{\bar{w}}_{\text{pred}} -\hat{\bar{w}}_{\text{ref}}\bigr\rVert_{\mathcal{B}}} {\bigl\lVert\hat{\bar{w}}_{\text{ref}}\bigr\rVert_{\mathcal{B}}}, 
    \qquad
    \lVert\hat{f}\rVert^2_{\mathcal{B}} = \sum_{\mathbf{k}\in\mathcal{B}} \lvert\hat{f}(\mathbf{k})\rvert^2.
        \label{eq:band_error}
\end{equation}
Figure~\ref{fig:spectral_dynamics} shows the resulting wavenumber-band error $\varepsilon_{\mathcal{B}}$ over the training iterations.
After $5{,}000$ Adam iterations, M5 has reduced the $A_0$-band error from $1.00$ to $0.97$, while M4 reaches $0.31$ and M6 $0.58$. The L-BFGS phase then reduces the $A_0$-band error by $76\,\%$ for M4, against $42\,\%$ and $10\,\%$ for the M6 and M5 baselines. Above the $A_0$ band, in the short-wavelength band $1.15\,k_{A_0}<|\mathbf{k}|\le2.0\,k_{A_0}$, the error of M5 increases during training and is not recovered. The $\tanh$ architecture therefore generates spatial content that the reference solution does not contain. M6 performs better, but only M4 reduces the relative error in the short-wavelength band below $0.2$.

\begin{figure}[htbp]
\centering
\includegraphics[width=1\linewidth]{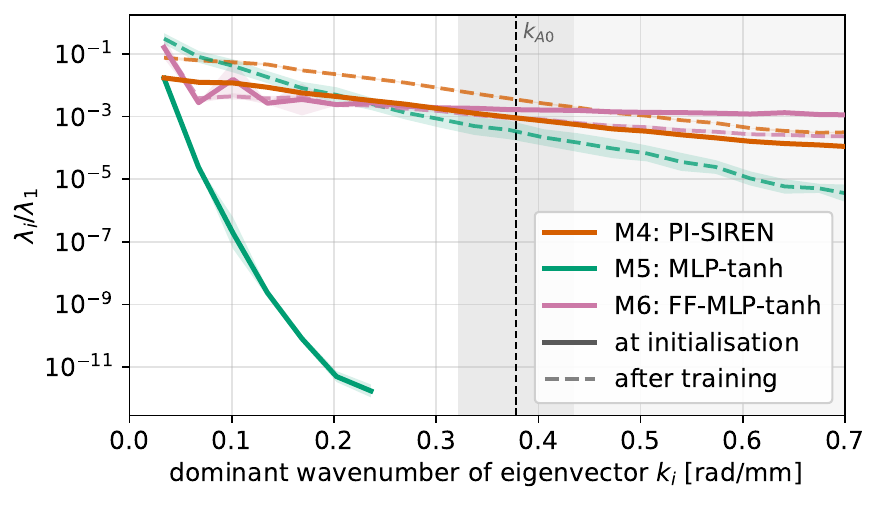}
\caption{NTK eigenvalue spectrum $\lambda_i/\lambda_1$ against the dominant wavenumber $k_i$ of each eigenvector at initialization and after training. The dashed line marks $k_{A_0}=2\pi/\lambda_{A_0}\approx0.38\,\mathrm{rad/mm}$.}
\label{fig:ntk_analysis}
\end{figure}

To separate the bias of the network architecture from the influence of optimization and training, the neural tangent kernel (NTK)~\cite{Jacot.2018} is computed for each architecture on a $30\times30$ uniform test grid in FP64. The NTK is the Gram matrix of the parameter gradients of the network output at the query points, assembled from the real and imaginary parts. It measures how sensitively the predicted wavefield responds to a parameter update at each spatial location. Diagonalizing it yields eigenpairs $(\lambda_i,\boldsymbol{\phi}_i)$ whose eigenvectors are spatial modes of the network output and whose eigenvalues act as the convergence rates of those modes. Under gradient flow, the error component along eigenvector $i$ decays as $\exp(-\lambda_i t)$, so a steep decay of the spectrum with wavenumber implies that high-wavenumber content is learned only very slowly~\cite{Rahaman.2018, Tancik.2020}. Figure~\ref{fig:ntk_analysis} illustrates $\lambda_i/\lambda_1$ against the dominant wavenumber $k_i$ at network initialization and after training. Each spectrum is normalized by its own largest eigenvalue. At initialization the plain MLP-$\tanh$ spectrum has collapsed to $\lesssim10^{-11}$ by $k_i\approx0.2\,\mathrm{rad/mm}$, below the $A_0$ band, in which no eigenmode exceeds $10^{-12}\lambda_1$. PI-SIREN M4 and the Fourier-embedded baseline M6 both retain $\lambda_i/\lambda_1\approx10^{-3}$ at $k_{A_0}$, a gap of approximately nine orders of magnitude. Because this is a comparison at initialization, it is independent of the architecture-specific training parameter and isolates the contribution of the network architecture to spectral bias.  
Comparing the SIREN and Fourier-embedded baseline results at $k_{A_0}$ reveals that the remaining performance gap ($L_2$ of $0.103$ versus $0.245$) is not due to a lack of sensitivity in the $A_0$ band. Rather, it is due to the kernel shape beyond this band. The Fourier-embedded kernel remains flat in the short wavelength band, while the PI-SIREN kernel decays. The flat tail of the Fourier-embedded kernel is consistent with its residual short-wavelength error in Figure~\ref{fig:spectral_dynamics} (b), while the collapsed M5 kernel explains why that network reduces neither band. Only the fully periodic SIREN drives the $A_0$ band to convergence, while the Fourier-embedded $\tanh$ MLP retains a higher band error. The Fourier embedding of M6 acts on the input only. It removes the spectral bias of the $\tanh$ network but fixes a broadband, approximately stationary kernel with a bandwidth set by $\sigma$. In the SIREN architecture, every layer applies a periodic activation function. The resulting kernel is able to cover the $A_0$ band, while also decaying in short-wavelength bands. 

\subsection{Reconstruction and Damage Identification on Experimental Data}
\label{sec:res_exp}
The proposed framework is applied to experimental wavefield measurements acquired with the optical microphone setup described in the ``Experimental Dataset'' section. The dominant response frequency is determined from the spectral peak of the full scan at $96.4\,\mathrm{kHz}$. The sensor density study performed above established that reliable damage localization requires sampling at or above the Nyquist limit ($\approx 2$~ppw). For the experimental data, this density is achieved with 625 points obtained through stratified subsampling of a dense scan rather than a structured grid. The structured grids used in synthetic studies assume noise-free samples, whereas experimental measurements carry optical and coupling noise. A rigid grid may place points at locations of locally degraded signal quality without spatial redundancy. Stratified sampling distributes points across spatial cells while selecting the most reliable measurement per cell, improving robustness at equivalent nominal density.

\begin{figure*}[t]
\centering
\includegraphics[width=\linewidth]{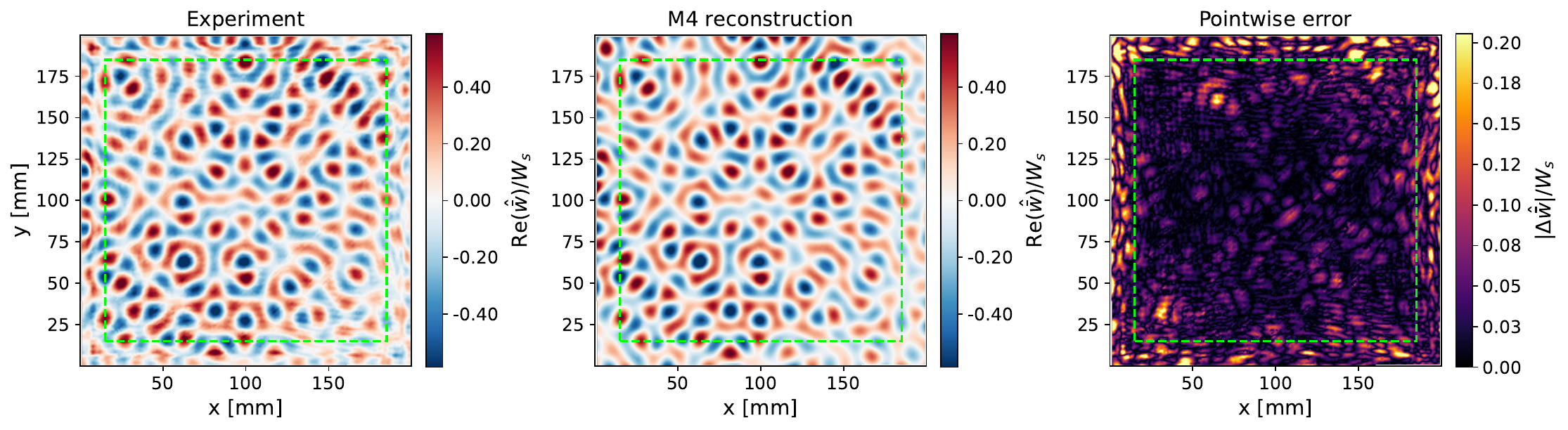}
\caption{Experimental pristine reconstruction: Experimental measurement (left), M4 PI-SIREN prediction (center), pointwise error $|\Delta \hat{\bar{w}}|/W_s$ (right). The dashed box indicates the interior region ($>15\,\mathrm{mm}$) used for the interior $L_2$ calculation.}
\label{fig:exp_recon}
\end{figure*}

The network architecture and three-phase optimization strategy (Adam, RAR, L-BFGS) follow Algorithm~\ref{alg:FreqDomain_PINN}. The boundary-condition weight is reduced to $\lambda_{\text{BC}} = 1$ because the idealized free-edge condition is only approximately satisfied by the physically supported plate. The RAR hard-point fraction is reduced from 50\% to 10\%, since the highest-residual collocation points concentrate on the free-edge boundaries, where the model mismatch dominates (cf. Figure~\ref{fig:exp_recon}). The remaining 90\% random points preserve uniform domain coverage. No synthetic noise is added during the Adam phase: the measurement noise in the experimental data already provides sufficient stochastic regularization. Figure~\ref{fig:exp_recon} illustrates the comparison of measurement, the PI-SIREN prediction and the pointwise error. The model reproduces the dominant $A_0$-mode interference pattern in the plate interior, including the radial wavefront from the central actuator. Reconstruction error concentrates near the boundaries (Figure~\ref{fig:exp_recon}, right): the global error is $L_2(\hat{\bar{w}}) = 29.4\,\%$, while over the interior region ($>15\,\mathrm{mm}$ from the edge) it decreases to $18.5\,\%$ (Table~\ref{tab:exp_l2}). Both values exceed the synthetic result at the same density ($10.3\,\%$ for $25^2$, Table~\ref{tab:ppw_study}). This increase is attributable to (i)~optical measurement noise, (ii)~boundary-condition differences between the idealized free-edge model and the physically supported plate, which dominate the edge region and (iii)~model-form error, as the single-frequency MR residual does not capture all physical features of the experimentally measured field. However, the physics-informed loss provides effective regularization on this noisy data, consistent with the dominant regularization effect observed in the performed synthetic ablation study.

\begin{table}[htbp]
\centering
\caption{Experimental pristine reconstruction error.}
\label{tab:exp_l2}
\begin{tabular}{lc}
\hline
Region & Relative $L_2(\hat{\bar{w}})$ \\
\hline
Full plate            & $29.4\,\%$ \\
Interior ($>15$~mm)   & $18.5\,\%$ \\
\hline
\end{tabular}
\end{table}
The trained surrogate serves as the pristine baseline in the damage identification pipeline described in Figure~\ref{fig:Damage_identification_pipeline}. The specimen contains a milled notch on the back side of the plate with dimensions $8 \times 2 \times 1.6\,\mathrm{mm}$ (length $\times$ depth $\times$ in-plane thickness), centered at $(150,150)\,\mathrm{mm}$. The damaged wavefield is measured on the same plate and the pristine prediction is aligned to it using the complex scalar $\alpha$ (Eq.~\eqref{eq:complex_alignment}), obtained via least-squares projection of the damaged field onto the pristine prediction. Here $|\alpha| = 0.87$ and $\arg(\alpha) = -11.6^\circ$. The scalar $\alpha$ reconciles global amplitude scaling and phase reference between the PI-SIREN baseline and optical measurement prior to subtraction.

\begin{figure*}[htbp]
\centering
\includegraphics[width=\linewidth]{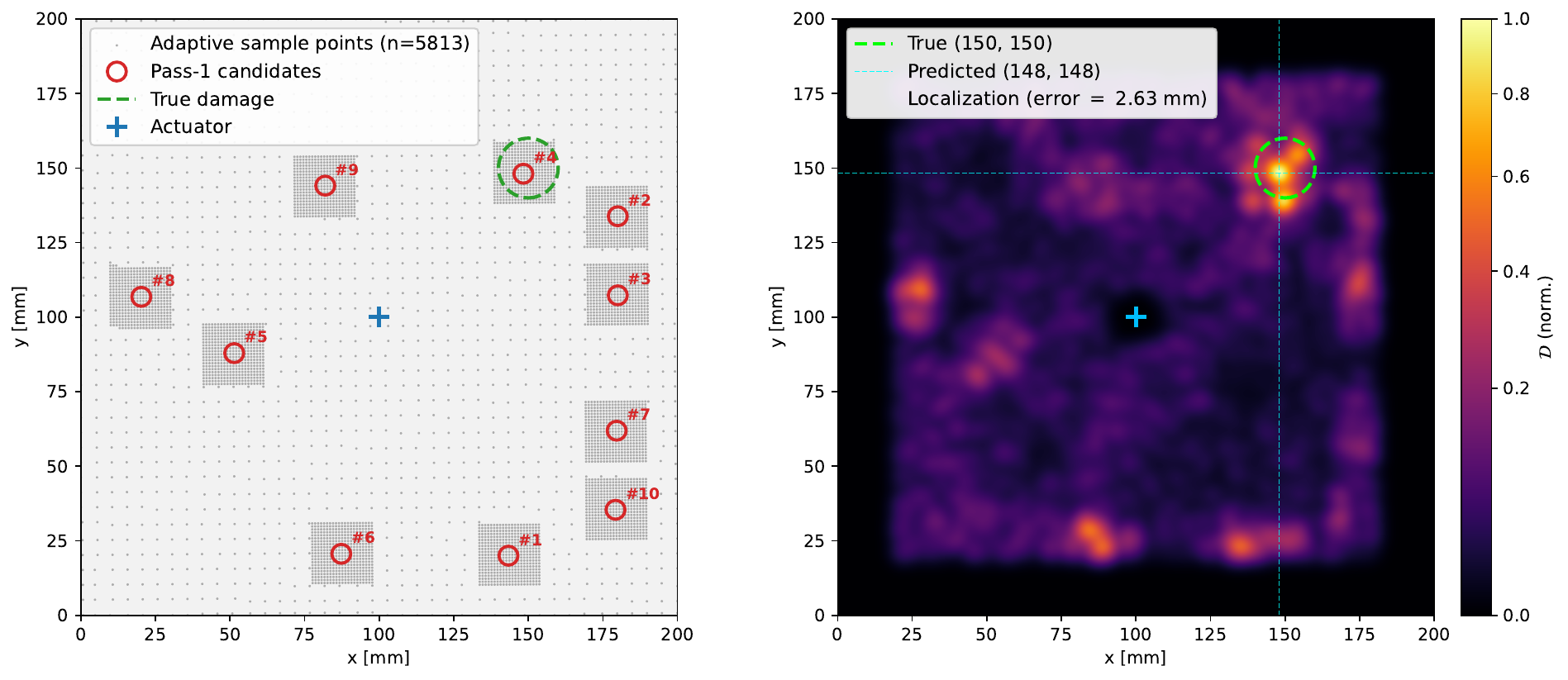}
\caption{Demonstration of damage identification pipeline on experimental data. Left: pass-1 candidate regions (\textcolor{red}{$\circ$}, \#1--10) and adaptive resampled distribution. The true damage is among the identified regions. Right: Damage indicator $\mathcal{D}$ recomputed on the adaptively refined sampling, with the true-defect region showing the strongest response ($2.63$~mm localization error).}
\label{fig:exp_method}
\end{figure*}

The scattered field $\hat{w}_{\text{scat}} = \hat{w}_{\text{dam}} - \alpha\,\hat{w}_{\text{pri}}$ in Eq.~\eqref{eq:scattered_field} is interpolated onto a regular grid, Gaussian-smoothed ($\sigma = 1.5\,\mathrm{mm}$) and mapped through the indicator $\mathcal{D} = \bigl|\mathrm{Im}\{ \overline{W_\sigma}\,\nabla^2 W_\sigma\}\bigr|$, which responds to localized scattering introduced by the defect. The proposed sequential pipeline (cf. Figure~\ref{fig:Damage_identification_pipeline}) is applied to the experimental data as illustrated in Figure~\ref{fig:exp_method}. A coarse pass over the full grid identifies candidate regions, after which an adaptive refinement pass densely resamples all candidate regions and recomputes the indicator. The true-defect region emerges as the dominant response. The refined localization places the damage centroid at $(148,148)\,\mathrm{mm}$, corresponding to a distance of $2.63\,\mathrm{mm}$ from the true position ($\approx 1\,\%$ of the plate diagonal). Localization sensitivity to post-processing choices is assessed by varying the smoothing width $\sigma$ over $[1.0,\,3.5]\,\mathrm{mm}$. The localization error varies by less than $0.35\,\mathrm{mm}$ across this range ($2.51$--$2.83\,\mathrm{mm}$, decreasing weakly with increasing $\sigma$), confirming that the result is robust to the kernel-width choice.

Damage localization succeeds ($2.63\,\mathrm{mm}$) despite the reconstruction error being above the synthetic threshold ($L_2 \lesssim 0.10$): the indicator responds to localized scattering, whereas reconstruction error is dominated by smooth, global mismatch that the $\alpha$-alignment and subtraction largely cancel. The task therefore tolerates larger reconstruction error on noisy experimental data than the noise-free synthetic threshold implies and that threshold should be regarded as a conservative design guideline. The adaptive, two-stage evaluation method restricts dense measurements to candidate regions identified by the coarse pass. This reduces the inspection scan from a full-plate, dense grid to approximately one-seventh of the measurement points. Combined with the inference-based pristine baseline, the complete inspection cycle therefore avoids the dense scanning that conventional baseline-subtraction methods require for both the reference and the damaged state. 

\subsection{Computational Cost}
\label{subsec:res_cost}
The framework separates computation into a one-time offline training phase and an online inspection phase (inference). To leverage accelerated automatic differentiation and XLA compilation, all model variants and ablation studies are implemented using JAX~\cite{bradbury2018jax} library. All models and ablation variants are trained on a compute cluster containing a single NVIDIA A100 GPU ($40\,\mathrm{GB}$ VRAM). 
\begin{table}[htbp]
\centering
\caption{Computational cost overview. Values for M4 trained on synthetic (FEM) and experimental (EXP) data.}
\label{tab:cost}
\begin{tabular}{lcc}
\hline
Quantity & FEM & EXP \\
\hline
M4 trainable parameters    & \multicolumn{2}{c}{$265{,}478$} \\
Training (one-time)        & $89.6$ min & $96.1$ min \\
\quad Adam / L-BFGS [min]  & $22.6 / 66.9$ & $18.7 / 77.5$ \\
Inference ($40{,}000$ pts) & $0.77$ ms & $0.76$ ms \\
\hline
\end{tabular}
\end{table}
As illustrated in Table~\ref{tab:cost}, training time varies from $90$--$96$~min, dominated by the double-precision L-BFGS refinement. Once trained, the frozen model evaluates $40{,}000$ query points in $< 1$~ms, enabling baseline reconstruction by inference rather than physical measurement. This offline-online split is practical because the dominant cost in wavefield-based inspection is measurement, not computation. At a robotic acquisition rate of $0.1$~s per point (travel, settling and measurement), a dense baseline scan of $40{,}000$ points requires approx. $1.1$~h of robotic acquisition, while the sparse scan used here ($625$ points) takes approx. $1$~min. The two-stage adaptive damage identification pipeline applies the same principle at inspection, densifying only candidate regions rather than resampling the full plate.

\section{Conclusion}
\label{sec:conclusion}
This work introduces a frequency-domain PI-SIREN framework for UGW inspection in isotropic thin-walled structures. By embedding the coupled MR plate equations directly into a SIREN-based model architecture, the method enables high-fidelity reconstruction of steady-state wavefields without temporal discretization. The sinusoidal activation functions allow the network to represent the high-frequency content that limits conventional MLP-based PINNs with $\tanh$ activation functions at high ultrasonic frequencies. The hybrid optimization strategy combining first-order Adam, RAR and quasi-Newton L-BFGS fine-tuning drives the reconstruction to low errors. Numerical ablation studies on synthetic FE data established that the full optimization strategy (Adam, RAR, L-BFGS) reduces the reconstruction $L_2$ error by $79\,\%$ relative to a Data-only baseline and by $23\,\%$ relative to the Adam + L-BFGS variant with uniform collocation. The RAR phase concentrates collocation points in the high-residual regions prior to quasi-Newton L-BFGS refinement. A sensor density study revealed that damage localization becomes reliable at the Nyquist sampling limit ($2$~ppw), while full-field reconstruction accuracy continues to improve until approximately $2.8$~ppw.

On experimental data acquired with a robotic optical microphone setup, the trained PI-SIREN surrogate serves as a robust pristine baseline despite higher global reconstruction error ($L_2 = 29.4\,\%$ versus interior $L_2 = 18.5\,\%$) compared to the noise-free synthetic case at equivalent density. Although the physics-informed loss effectively regularized noisy measurements within the domain, the residual error is concentrated near the boundary edges. The implemented idealized free-edge boundary condition only approximates a physically supported plate, resulting in a less reliable reconstruction of the experimental data. Integrated into the two-stage damage identification pipeline (coarse candidate detection followed by adaptive resampling), the model localizes a back-side milled notch with an absolute error of $2.63\,\mathrm{mm}$ ($\approx 1\,\%$ of the plate diagonal).

The proposed framework addresses three key limitations of existing SHM/NDT approaches: (i)~the computational cost of FE-based baselines for repeated evaluations (ii)~the data requirements and lack of interpretability of purely data-driven models and (iii)~the spectral bias of conventional MLP-$\tanh$-based PINNs at ultrasonic frequencies. This is achieved through periodic activations of the SIREN architecture~\cite{Sitzmann.2020}. The present study demonstrates the framework for a single plate configuration, material and excitation frequency. The practical advantage scales with the number of configurations covered. The one-time training cost is amortized across all subsequent inspections of the same or related structures. Operator-based architectures such as DeepONets~\cite{Lu.2021, Goswami.2023} represent a promising approach towards a single multi-configuration surrogate spanning multiple materials and geometries. Another approach is to directly utilize the PINN as inverse model for defect parameters. Rather than subtracting a pristine baseline, the damage location and extent would be included in the physics residual as trainable quantities. While this approach eliminates the need for pristine measurements of each specimen, it requires retraining for every new damaged dataset, shifting the cost from acquisition to computation. Further research involves applying the model framework to anisotropic and composite plate structures.

\section*{Acknowledgments}
\footnotesize
The authors gratefully acknowledge the computational and data resources provided through the joint high-performance data analytics (HPDA) project ``terrabyte'' of the German Aerospace Center (DLR) and the Leibniz Supercomputing Center (LRZ). 
During the preparation of this publication, Qwen3.8-27B and DeepL Write were utilized for language and grammar refinement and assistance in research data visualization. After utilizing these tools, the authors thoroughly reviewed and edited the content and accept full responsibility for the accuracy and integrity of the final publication.

\section*{ORCID iDs}
\footnotesize
\raggedright\urlstyle{same}\ignorespaces
Lukas Jilke \orcidlink{0000-0001-6879-3279} \url{https://orcid.org/0000-0001-6879-3279}\\
Caspar Wasle \orcidlink{0009-0003-7198-9186} \url{https://orcid.org/0009-0003-7198-9186}\\
Norbert Hosters \orcidlink{0000-0003-1174-4446} \url{https://orcid.org/0000-0003-1174-4446}\\
Florian Raddatz \orcidlink{0000-0002-0660-7650} \url{https://orcid.org/0000-0002-0660-7650}\\
Marek Behr \orcidlink{0000-0003-4257-8276} \url{https://orcid.org/0000-0003-4257-8276}\\
Gerko Wende \orcidlink{0000-0003-2368-2369} \url{https://orcid.org/0000-0003-2368-2369}\\

\section*{CRediT authorship contribution}
\footnotesize
\textbf{L.J.} performed the roles of Conceptualization, Data curation, Formal analysis, Investigation, Methodology, Project administration, Software, Validation, Visualization, Funding acquisition, Writing -- original draft, Writing -- review \& editing. \textbf{C.W.} performed the roles of Investigation, Writing -- review \& editing. \textbf{N.H.} performed the roles of Conceptualization, Supervision, Writing -- review \& editing. \textbf{F.R.} performed the roles of Funding acquisition, Writing -- review \& editing. \textbf{M.B.} performed the roles of Supervision, Writing -- review \& editing. \textbf{G.W.} performed the roles of Supervision, Writing -- review \& editing. 

\section*{Funding}
\footnotesize
The author(s) received no financial support for the research, authorship, and/or publication of this article.

\bibliographystyle{ieeetrdoi}
\bibliography{references}  

\end{document}